\documentclass[11pt]{article}
\usepackage{ tikz, color, caption, ulem}
\usetikzlibrary{decorations.markings,arrows,calc, shapes}
\definecolor{darkblue}{rgb}{0, 0, 0.3}
\usepackage[colorlinks=true, pdfstartview=FitV, linkcolor=darkblue, citecolor=darkblue, urlcolor=darkblue]{hyperref}
\usepackage{cancel, mathrsfs, amsmath, amssymb, todonotes}

\usepackage{geometry}
\def\CV{{\mathcal M}}
\def\CVh{\widehat{\CV}}
\def\zetat{\tilde{\zeta}}
\def\zt{\tilde{z}}
\def\CM{{\mathbb G}}
\def\ver{r}
\def\verB{R}
\def\a\alpha
\def\bs{\boldsymbol}
\def\le{\left}

\def\ri{\right}
\def \eqref#1{(\ref{#1})}
\def \bs{\boldsymbol}

\def\br{\begin{remark}}
\def\er{\end{remark}}

\def\bl{\begin{lemma}}
\def\el{\end{lemma}}
\def\bd{\begin{definition}}
\def\ed{\end{definition}}
\def\bp{\begin{proposition}}
\def\ep{\end{proposition}}
\def\be#1\ee{\begin{align}#1 \end{align}}

\def\Jum{J}
\def\s{\sigma}
\def\kat{\tilde{\kappa}}
\def \pa{\partial}
\def\a{\alpha}
\def\Ec{{\mathcal E}}
\def\Mcal {{\mathcal M}}
\def \ds {\displaystyle}
\def\1{\mathbf 1}
\def\wt{\widetilde}

\def \wh {\widehat}

\definecolor{shadecolor}{rgb}{0.95, 0.95, 0.86}
\def\g{\gamma}
\def \la{\label}

\def \DD{\mathbb D}
\def \QED {\hfill $\blacksquare$\par \vskip 3pt}

\def\J{\mathbb J}
\def\M{\mathcal M}

\makeatletter
\@addtoreset{equation}{section}
\makeatother

\newtheorem{theorem}{Theorem}[section]
\newtheorem{example}{Example}[section]

\newtheorem{lemma}{Lemma}[section]
\newtheorem{remark}{Remark}[section]
\newtheorem{proposition}{Proposition}[section]
\newtheorem{corollary}{Corollary}[section]
\newtheorem{definition}{Definition}[section]
\newtheorem{conjecture}{Conjecture}

\def\bt{\begin{shaded}\begin{theorem}}
\def\et{\end{theorem}\end{shaded}}
\def\bea#1\eea{\begin{align}#1\end{align}}
\def\beas{\begin{eqnarray*}}
\def\eeas{\end{eqnarray*}}
\def \pa{\partial}
\def\L{\mathcal L}
\def\N{{\mathbb N}}

\def \d{\,\mathrm d}
\def\1{{\bf 1}}
\def \pa{\partial}

\def\nn{\nonumber}
\def\ba{\begin{array}}
\def\ea{\end{array}}
\def\la{\label}
\def\p{\partial}

\def\C{{\mathbb C}}
\def\R{{\mathbb R}}

\def\gt{{\widetilde {\gamma}}}
\def\Ccal{\mathcal C}
\def\f{\frac}
\def\Oh{
\mathcal W
}
\def\ka{\kappa}
\def\tr{{\rm tr}}
\begin{document}

\vspace{0.2cm}
\begin{center}
\begin{Large}
{ Extended Goldman symplectic structure in Fock-Goncharov coordinates}
\end{Large}\\
\bigskip
M. Bertola$^{\dagger\ddagger}$\footnote{Marco.Bertola@\{concordia.ca, sissa.it\}},  
D. Korotkin$^{\dagger}$ \footnote{Dmitry.Korotkin@concordia.ca},
\\
\bigskip
\begin{small}
$^{\dagger}$ {\it   Department of Mathematics and
Statistics, Concordia University\\ 1455 de Maisonneuve W., Montr\'eal, Qu\'ebec,
Canada H3G 1M8} \\
\smallskip
$^{\ddagger}$ {\it  SISSA/ISAS,  Area of Mathematics\\ via Bonomea 265, 34136 Trieste, Italy }\\
\end{small}
\vspace{0.5cm}
\end{center}

 \begin{center} \bf Abstract \end{center}
 {The goal of this paper is to express the extended Goldman symplectic structure on the $SL(n)$ character variety of a punctured Riemann surface in terms of Fock-Goncharov coordinates. The associated symplectic form has integer coefficients expressed via the inverse of the Cartan matrix. The main technical tool is    a canonical  two-form associated to a  flat graph connection. 
We discuss the relationship between the extension of the Goldman Poisson structure and the Poisson structure defined by Fock and Goncharov. 
We elucidate the role of the Rogers' dilogarithm as generating function of the symplectomorphism defined by a graph transformation.}

\tableofcontents

\section{Introduction}

The 
$SL(n)$ character variety  of a Riemann surface with $N$ punctures  and negative Euler characteristics is equipped with 
the canonical Goldman Poisson bracket \cite{Goldman}  (see p.266 of
\cite{Goldman1}):
\be
\Big\{\tr M_\g,\;\tr M_\gt\Big\}_G= \sum_{p\in \g\cap \gt} \nu(p)\,\left( \tr (M_{\g_p\gt})
-\f{1}{n}  \tr M_\g  \tr M_\gt  \right)
\la{Goldmanbr}
\ee
for any two elements $\g\;,\gt\in \pi_1(\Ccal)$, where $\nu(p)=\pm 1$ is the contribution of point $p$ to the intersection index of $\g$ and $\gt$.
 For $N\geq 1$ the Goldman  bracket  (\ref{Goldmanbr}) is degenerate, with the Casimirs being the spectral invariants  of the  local monodromies around the
 punctures.

The Goldman Poisson structure is the canonical Poisson structure on the moduli space of flat connections; it is implied by the canonical Atiyah-Bott Poisson structure on the space of $SL(n)$ connections over the punctured Riemann surface (with an appropriate condition on the singularity structure near the 
punctures).
 The expression for the symplectic form which inverts the Goldman bracket  \cite{Goldman1} on  symplectic leaves was found by Alekseev and Malkin \cite{AM}.
 The form of \cite{AM} admits a natural nondegenerate extension;
 under such an extension the space is augmented by natural canonical partners to the Casimir functions. These extended  spaces 
were introduced  in \cite{Jeffrey} where the symplectic forms on them were induced from extended spaces of flat connections and then reprised and generalized in  \cite{Boalch2}.

The goal of this paper is to provide  an explicit expression of the extended Goldman's symplectic form using Fock-Goncharov coordinates \cite{FG} and show that these coordinates are  log-canonical  for the Poisson structure, which means  that the Poisson brackets of the logarithms of any two coordinates is a constant. The main motivation comes from the idea of interpretation of the isomonodromic tau-function of Miwa and Jimbo as generating function of the monodromy
symplectomorphism \cite{BKtau} which requires an explicit construction of the symplectic potential of the extended Goldman symplectic form and study of its transformation properties.

To present our results in more detail we  introduce  a set of generators of $\pi_1(\Ccal\setminus\{t_j\}_{j=1}^N,\,z_0)$  which satisfy the relation
 \be
 \g_1\dots,\g_{_N}\prod_{j=1}^g \a_j\beta_j^{-1}\a_j^{-1}\beta_j=id\;.
 \ee
{We use a slightly non-standard  relation between the generators in order   to follow the conventions of  \cite{AM} and facilitate the comparison.}
 Given a monodromy representation $\pi_1(\mathcal C\setminus\{t_j\}_{j=1}^N, z_0) \to SL(n)$, the corresponding monodromy matrices satisfy the  same relation
 \be
 M_{\g_1}\dots M_{\g_{_N}}\prod_{j=1}^g A_j B_j^{-1} A_j^{-1} B_j=\1\;.
 \la{relfun}
 \ee
 We are going to consider the subspace $\CV$ of the character variety such that all monodromies $M_k=M_{\g_k}$ are  diagonalizable
 \be
 M_{\g_k}=C_k \Lambda_k C_k^{-1}
 \ee
 where $\Lambda_k$ are diagonal matrices with distinct eigenvalues; these are the Casimirs of the Goldman Poisson structure.
 On a symplectic leaf  $\CV_{\Lambda}$ the Goldman's bracket is invertible and  the symplectic form is given by
 \cite{AM}:
\bea
\la{OMAM}
 \Oh_G=   
 \frac 1 2 \bigg(&\sum_{i=1}^{2g+N}\tr\Big(  \d K_{i-1} K_{i-1}^{-1} \wedge   \d K_{i} K_{i}^{-1} 
\Big)
 + 
\sum_{j=1}^{N} \tr \Big( \Lambda_j C_j^{-1}\d C_j \wedge \Lambda_j^{-1} C_j^{-1}\d C_j\Big)
\\
&\nn + 2\sum_{\ell=1}^{2g}\tr \Big(   P_{\ell}^{-1}\d P_{\ell}\wedge  D_\ell^{-1}\d D_\ell  \Big)
\nn
+  \sum_{\ell=1}^{2g}\tr \Big(  D_\ell^{}  P_{\ell}^{-1}\d P_{\ell} \wedge  D_\ell^{-1} P_{\ell}^{-1}\d P_{\ell}  \Big)  \bigg)
\eea
where
\bea
K_j  &  := M_1\cdots M_j\;, \qquad\hspace {1cm} j\leq N \;,\nn\\
\label{defK}
K_{N+2j-1}  & = K_{N+2j-2} A_j\;,\ \  K_{N+2j} = K_{N+2j-1} B_j^{-1} A_j^{-1} B_j\ , \ \\
\nn& j=1,\dots, g\;.
\eea
The  matrices $\Lambda_j$ are diagonal matrices belonging to the  Cartan torus $\mathfrak h $ of  $SL(n)$, and $C_j$ are matrices of eigenvectors of $M_j$,
 \be
 M_j = C_{j} \Lambda_j C_{j}^{-1}\; , \hspace{0.8cm}\ \ j=1,\dots, N
 \la{MC}\ee
 while the diagonal form of the matrices $A_\ell$ 
 is given by 
 \be
A_\ell =P_{2\ell-1} D_{2\ell-1} P_{2\ell-1}^{-1}
\la{diaA}
 \ee
 and $ B_\ell$ enters in the relation below
  \be
 P_{2\ell}:=B_\ell^{-1} P_{2\ell-1}\;, \hspace{0.8cm} D_{2\ell}:=D_{2\ell-1}^{-1}\;.
 \la{PBP}
 \ee
 The form \eqref{OMAM} is invariant under the following toric action:
 \be
 \nn
 &C_j\mapsto C_j H_j\; , \ \ \ &j=1,\dots, N;\\
 \nn 
 &P_{2\ell-1} \mapsto P_{2\ell-1} W_\ell\;, \ \ 
 P_{2\ell} \mapsto P_{2\ell} W_\ell\;, \ \qquad & \ell=1,\dots,g 
 \ee
 where $H_j, W_\ell$'s are arbitrary matrices in the Cartan torus of $SL(n)$.

 Following  \cite{Jeffrey} we introduce the extended space  $\CVh$  defined   by the following quotient of the  space of matrices $\{A_j, B_j\}_{j=1}^g,
 \{C_j,\Lambda_j\}_{j=1}^N$ with 
 $\Lambda_j$ being traceless diagonal matrices,
   satisfying one relation:
 \be
 \CVh:= \Bigg\{ &\{A_j, B_j\}_{j=1}^g\, ,
 \{C_j,\,\Lambda_j\}_{j=1}^N:
&C_1\Lambda_1  C_1^{-1}\dots C_{_N}\Lambda_N C_{_N}^{-1}  \prod_{j=1}^g A_j B_j^{-1} A_j^{-1} B_j=\1\Bigg\}\Big/\sim\;.
\la{defCVh}
 \ee
where $\sim$ means equivalence of  the sets of matrices differing by simultaneous transformation $M_\gamma\to G M_\gamma G^{-1}$, $C_j\to G C_j$ with any $G\in SL(n)$. 

The space $\CVh$ is a torus fibration over $\CV$ with a fiber consisting of $N$ copies of the Cartan torus.
On $\CVh$ we introduce the two-form
\be
 \Oh =\Oh_G+ \sum_{j=1}^N \tr\Big(  C_j^{-1}\d C_j\wedge \Lambda_j^{-1} \d\Lambda_j\Big)
\la{defCh}
\ee
where     $\Oh_G$ is given by (\ref{OMAM}).
We prove that   the form \eqref{defCh} has constant coefficients when expressed in terms of the logarithms of Fock--Goncharov coordinates (since our presentation here is local we don't discuss the problem of choice of the branches of logarithms; the symplectic form is invariant under this choice). 
Thus, on the  coordinate charts parametrized by the Fock-Goncharov coordinates the form is manifestly closed.  These coordinates are defined on a Zariski open (dense) set, and since the form \eqref{defCh} is  analytic, it  follows that it is closed on the whole $\CVh$. 

The Casimirs $m_{k;j} $   on the space 
$\CV_{\Lambda}$ are defined by
$$
\Lambda_k = {\bf m}_k^{\boldsymbol \a} =
{\rm diag}\le(m_{k;1} , \frac {m_{k;2}}{m_{k;1}}, \dots, \frac {m_{k;n-1}}{m_{k;n-2}}, \frac 1{m_{k;n-1}}\ri)\;.$$
The Casimirs $m_{k;j} $ are  certain monomials  of the Fock-Goncharov coordinates. 

The following theorem is the main result of this paper:
\begin{theorem}
Denote by $\sigma_{j}$ the logarithms of Fock-Goncharov coordinates associated to a given triangulation 
of $\Ccal$ with $N$ vertices. Then the form
$\Oh$ \eqref{defCh} can be expressed as follows
 \be\la{Ohs}
 \Oh= \sum_{j<k} n_{jk}    \d \sigma_j\wedge \d \sigma_k 
+
 n\sum_{k=1}^N \sum_{j=1}^{n-1}
  {\d \log m_{k;j} \wedge \d \rho_{k;j}}
 \ee
 where $\rho_{k;j}$, $j=1,\dots,N$, $k=1,\dots,n-1$ are the toric variables corresponding to the Casimirs  $m_{k;j} $, which are exponents of linear combinations of $\sigma_j$'s with integer coefficients.
The  integers $n_{jk}$ are computed explicitly in Section \ref{secsymp}. 
\end{theorem}
 
 The theorem implies in particular that the form $\Oh_G$ (\ref{OMAM}) restricted to the  symplectic leaf $\Lambda=const$ is given by
 $$
 \Oh_G\big|_{\Lambda=const}= \sum_{j<k} n_{jk}    \d \sigma_j\wedge \d \sigma_k \big|_{\Lambda=const}\;.
 $$
 Then the representation (\ref{Ohs})  implies that the form $\Oh$ is non-degenerate, and, therefore, symplectic.

 The proof  follows several steps; the main tool is a canonical two-form associated to  an arbitrary oriented graph $\Sigma$ embedded in a Riemann surface;    a ``jump" matrix $J_{e}=J(e)$ is  assigned to each edge $e$ of $\Sigma$ such that the monodromy  around each multi-valent vertex of $\Sigma$ is trivial. Enumerating the edges $e$ incident to each vertex  $v$ of $\Sigma$ in counterclockwise order
 starting from a chosen ``cilium" the form can be written as 
 \be
 \Omega(\Sigma) =\sum_{v\in {\bf V}(\Sigma)}  \sum_{e_1\prec e_2 \perp v} \tr \bigg(J_{e_1}^{-1}\d J_{e_1} \wedge J_{e_2}^{-1}\d J_{e_2}\bigg)
 \la{Omegaint}\ee
 where $e_1\prec e_2$ means that $e_1$ precedes $e_2$ in counterclockwise order under 
 the chosen ciliation.
 In this expression it is assumed that all edges incident to the vertex $v$ are oriented outwards and
 under the change of orientation we have $J(-e)=J^{-1}(e)$.
 We  show that  $\Omega(\Sigma)$  is invariant under certain transformations  of the graph. 
The  form $\Oh$ \eqref{Ohs}   can then be represented as   ${\f{1}{2}}\Omega(\Sigma_{AM})$ for  a suitable  graph $\Sigma_{AM}$ associated to the Alekseev-Malkin formalism \cite{AM}, see Section \ref{secAM}.  
 
 The second step is to  consider another  graph $\Sigma_{FG}$ with jump matrices defined in  Fock-Goncharov's formalism \cite{FG} and show that 
 the associated symplectic form $\Omega(\Sigma_{FG})$ has  log-canonical form with respect to Fock-Goncharov coordinates and the toric variables. 
Finally,  by a sequence of transformations we can transform the graph $\Sigma_{FG}$ to the  graph $\Sigma_{AM}$ 
 and get the equality $\Omega(\Sigma_{FG})=\Omega(\Sigma_{AM})$ 
 which leads to  (\ref{Ohs}).

 As it was shown in \cite{ZheSun,SZ}, the 
 Fock-Goncharov Poisson structure \cite{FG} is equivalent to the Goldman bracket on  symplectic leaves. From this fact one can deduce that  the Poisson brackets between the variables $\sigma_j$ implied by the form $\Oh$ (\ref{Ohs}) also coincide with the Fock-Goncharov brackets.
 On the other hand, the  Poisson brackets  between toric variables  $\rho_{k;j}$  and the brackets between the toric variables and variables $\sigma_j$ implied by the form $\Oh$ are found and proved here in
 $SL(2)$ and $SL(3)$ cases; for higher groups we formulate a conjecture based on computer experiments.

 We conclude the paper by considering the generating function of the symplectomorphism induced by the change of triangulation
in the $SL(2)$ case. 
We show that  this generating function  is given by  the Rogers' dilogarithm, similarly to the construction of the paper \cite{Gekhtman}. This leads to definition of the   dilogarithm line bundle over
a real component of the extended character variety.

 \begin{remark}\rm
 After completion of this work we were alerted about the extensive papers by Z. Sun, A. Wienhard and T. Zhang \cite{SWZ,SZ}. In these papers the authors address the problem of the computation of Darboux coordinates for the Goldman bracket on the Hitchin's component of the  $PGL(n)$ character variety  of an unpunctured Riemann surface of genus $g$.  In particular, the papers \cite{SWZ,SZ} contain Darboux coordinates for the Goldman symplectic form
 (Th.8.22 of \cite{SWZ}) which in the $PGL(2)$ case coincide with Fenchel-Nielsen coordinates, reproducing the classical result of Wolpert \cite{W1,W2}. The direct comparison with results of
  this paper is not straightforward since here we are dealing with (the extension of) the Goldman Poisson bracket and 
  symplectic form on the character variety of punctured Riemann surface; our formalism does not cover the 
  un--punctured case. On the other hand, the formalism of \cite{SWZ,SZ} was not explicitly extended to the
  punctured case (in particular, in the $2\times 2$ case the Fenchel-Nielsen coordinates are related to shear coordinates in a highly non-trivial way). Therefore, the problem of finding an explicit relationship between   an extension of \cite{SWZ,SZ} to
  the punctured case and  results of this work remains open.
  \end{remark}

\noindent {\bf A warning on conventions.}
Given a symplectic manifold $(\mathcal M, \omega)$ we follow the convention (used for example in \cite{BBT}) that the  Poisson bracket is constructed as follows:
we write $\omega = \sum_{i<j} B_{ij} \d x_i \wedge \d x_j$ in local coordinates.  Then the corresponding  Poisson bracket is  $\{x_i,x_j\} = P_{ij}$ with $P$ the {\bf inverse transposed} matrix to $B$ (in the other standard sources like the textbook by Arnol'd the convention would be that $P$ is  the inverse of $B$, thus leading to a relative difference of a sign). 
For example, in two dimensions 
\be
\{p,q\}=1    \ \Leftrightarrow\ \ \omega = \d p\wedge \d q\; .
\ee
With this convention for the Poisson  tensor $\mathbb P= \mathbb \pa_p \wedge \pa_q$ we have
 $\mathbb P \circ \omega=-{\rm Id}$ and $PB=-I$.

 \section{  Graphs on surfaces  and the canonical two-form}
 \la{secinv}
 
Let $\Sigma$ be a  finite embedded graph on a surface $\mathcal C$. 
We shall assume that the faces of the graph are simply connected although most of results of this section 
hold without this assumption.
Denote by ${\bf V}(\Sigma)$ the set of vertices and by ${\bf F}(\Sigma)$  the set of faces of $\Sigma$.
We 
use the notation ${\bf E}(\Sigma)$ for the set of edges of $\Sigma$ with all possible orientations.
Namely, for an edge connecting two vertices, $v_1$ and $v_2$ of $\Sigma$, we denote by $e$ the
oriented edge $[v_1,v_2]$ and by $-e$ the oriented edge $[v_2,v_1]$. Both $e$ and $-e$ are elements of
 ${\bf E}(\Sigma)$. 
 
 We shall use the following  notation: an (oriented) edge $e$ is called {\it incident} at  the vertex $v$, denoted by the symbol $e\perp v$,  if the corresponding unoriented edge  is incident in the usual sense and furthermore the orientation of $e$ has $v$ as source. Similarly, a face $f$ is called {\it incident} at  the vertex $v$, denoted by the symbol $f\perp e$, if $v\in \p f$.

 \bd
The pair $(\Sigma, J)$ consisting of an  oriented  graph $\Sigma$, considered up to isotopy, on a   surface $\Ccal$ of genus $g$ with $N$ punctures $\{t_1,\dots, t_N\}$ and a map $J:{\bf E}(\Sigma) \to SL(n)$ is called {\rm canonical}  if it satisfies the following conditions:
 \begin{enumerate}
 \item
 The only univalent vertices of $\Sigma$ are at the punctures $t_j$, $j=1,\dots,N$. For each of them there is a small  disk $\DD_j$ bounded by a   loop $s_j$ starting and ending at a vertex $q_j$ on the edge incident at $t_j$ and traversed counterclockwise. The disks and the bounding loops $s_j$  are required to be pairwise disjoint. 
\item The map $J:{\bf E}(\Sigma)\to  SL(n)$ has the property that $J(-e) = J(e)^{-1}$.
\item For each vertex $v\in {\bf V}(\Sigma)$ of valence $n_v\geq 2$, denote by  $\{e_1,\dots, e_{n_v}\}$  the   incident edges, oriented away from $v$ and enumerated in counterclockwise order starting from an arbitrary edge. Then we require the triviality of the total monodromy around the vertex:
\be
J(e_1)\cdots J(e_{n_v}) = \1\;.
\label{nomono}
\ee
 \end{enumerate}
 \ed
\noindent {\bf Flat connection on the dual graph.}
The above definition (and the use of the term ``jump'' for the matrix $J$) is motivated by the theory of Riemann--Hilbert problems.
An equivalent formulation can be given in terms of a {\it connection} on the dual graph. 

To be more precise, let us 
introduce the  dual oriented graph $ \Sigma^*$; vertices of $\Sigma^*$ are in one-to-one correspondence with faces (connected regions of the complement) of $\Sigma$, and faces of $\Sigma^*$ are in one-to-one  correspondence with the vertices of $\Sigma$. 
Two vertices  of $\Sigma^*$ are connected by an edge $e^*$ if the corresponding faces of $\Sigma$ share an edge $e$. The orientation of $e^*$ is chosen so that the intersection number of 
$e$ and $e^*$ equals to $1$.

The matrices $J$ can then be interpreted as  connection matrices along the edges  of $\Sigma^*$; the parallel transport between two vertices of $\Sigma^*$ along a path consisting of several  edges is  the product of the matrices $J$  of the  edges traversed in the given orientation.

The  condition \eqref{nomono} of triviality of monodromy around each non-univalent vertex of $\Sigma$  can be alternatively formulated as follows:  any closed loop in   $\Sigma^*$ which is trivial in $\pi_1(\Ccal\setminus \{t_j\}_{j=1}^N)$, has trivial holonomy, or, equivalently, that the connection on the graph $\Sigma^*$ is 
{\it flat}.

\noindent {\bf The canonical two-form.} To each canonical  pair $(\Sigma,J)$  we associate the following {\it canonical} two-form (we omit explicit reference to $J$ in the notation): 
 \be
 \Omega(\Sigma)= \sum_{v\in {\bf V}(\Sigma) }\sum_{\ell=1}^{n_v-1} \tr \bigg(  (\Jum^{(v)}_{[1:\ell]})^{-1} \d \Jum^{(v)}_{[1:\ell]} \wedge (\Jum_{\ell}^{(v)})^{-1}\d \Jum_{\ell}^{(v)}\bigg)
 \la{dThgen}
 \ee
 where ${\bf V}(\Sigma)$ denotes the set of vertices of $\Sigma$, $n_v$ is the valence of the vertex $v$ and $J^{(v)}_{\ell}$ are the jump matrices associated to the edges $e_1,\dots,e_{n_v}$ incident at $v$ (hence oriented outwards, according to our established usage of {\it incidence}), and enumerated in counterclockwise order and we use the following shorthand notation:
 $$
 \Jum^{(v)}_{[1:\ell]}=\Jum_{1}^{(v)}\dots\Jum_{\ell}^{(v)}\;.
 $$
 We observe that the expression  (\ref{dThgen})  is invariant under cyclic reordering of the edges thanks to \eqref{nomono}. Being written as the double sum, the expression (\ref{dThgen}) coincides with (\ref{Omegaint}).
 
 The form $\Omega(\Sigma)$ can be shown to be closed.  Namely, according to   \cite{AlMalMein}, for any set of  $s$ matrices $J_1,\dots, J_s$ the exterior derivative of  the form 
 \be
 \omega:= \sum_{\ell= 1}^s  \tr \le(K_{\ell}^{-1} \d K_{\ell} \wedge  J_\ell^{-1} \d J_\ell \ri) ,\qquad K_\ell:= J_{1} \cdots J_{\ell}
 \la{defo}
 \ee
is given by  $\d \omega = \frac 1 {12} \tr \big(\mu^{-1} \d \mu \wedge \mu^{-1} \d \mu \wedge \mu^{-1} \d \mu \big)$, with $\mu = J_1\cdots J_s$. Therefore the form $\omega$ is closed on the level set  $\mu = \1$. The expression (\ref{dThgen}) in then  a direct sum of several copies of (\ref{defo}) (one for each vertex of $\Sigma$, or face of $\Sigma^*$), subject to constraints due to the fact that the matrix associated to each edge  appears in the form at  two vertices of $\Sigma$ (or bounds two faces of $\Sigma^*$). 

The main part of the formula for $\Omega(\Sigma)$ appeared in \cite{Bertocorr} as a result of a computation for the exterior derivative of the so--called Malgrange one-form associated to a Riemann--Hilbert problem. This also  explains the use of the term ``jump matrix'' for $J$. 

\subsection{Invariance of $\Omega(\Sigma)$ under standard moves}
\label{secmoves}
The form $\Omega(\Sigma)$  (\ref{dThgen}) enjoys  invariance properties under certain  transformations of the pair $(\Sigma, J)$ which we call  ``canonical moves''. They are explained in this section.

\begin{figure}[htb]
\begin{center}
\includegraphics[width=0.8\textwidth]{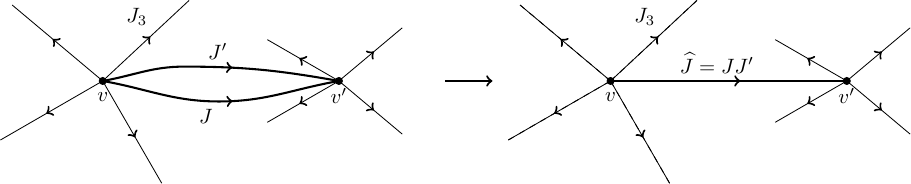}
\end{center}
\caption{Zipping two edges together.}
\label{zipping}
\end{figure}

\begin{lemma}[``Zipping'' lemma]\label{zipedge}
Suppose $e,  e'\in {\bf E}(\Sigma)$ have the same endpoints and are homotopic to each other. Let us  merge them together to an edge $\wh e$ and set $\wt J(\wh e) = J(e) J( e')$  (see Fig. \ref{zipping}) whereas $\wt J(e'') = J(e'')$ for any other edge $e''$. Denoting by $(\wt \Sigma, \wt J)$ the resulting  pair,  we have  $\Omega(\Sigma) = \Omega(\wt \Sigma)$.
\end{lemma}
\QED

{\bf Proof.}
Let $v,\wt v$ be the two vertices connected by $e,  e'$; consider first the case $v \neq  \wt v$. 
We assume that $e, e'$ are oriented away from $v$ and towards $\wt v$ and they are the first two edges in the  counterclockwise order at $v$. Then we can also enumerate the edges at $\wt v$ so that $ e'$ is the first edge and $e$ the second. 

\begin{figure}[htb]
\begin{center}
\includegraphics[width=0.7\textwidth]{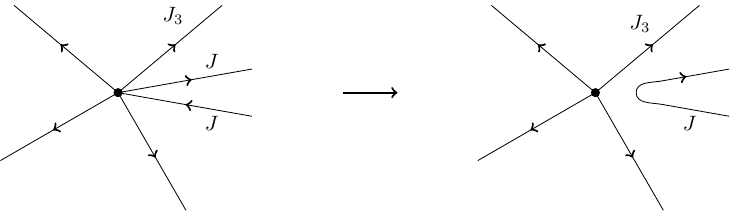}
\end{center}
\caption{Attaching and detaching an edge to/from a  vertex.}
\label{detach}
\end{figure}

Denote $J_1,\dots, J_{n_v}$ the jump matrices on the edges meeting  at $v$ and $\wt J_1,\dots \wt J_{n_{\wt v}}$ those  on the edges meeting at $\wt v $: under our convention  $J_1 = \wt J_2^{-1}$ and $J_2 = \wt J_1^{-1}$.   Consider now the affected contributions to the form $\Omega(\Sigma)$:
\be
 &\overbrace{\tr \Big( (J_1J_2)^{-1} \d (J_1J_2) \wedge J_2^{-1} \d J_2  \Big) + \dots  }^{\hbox{contribution at $v$}} 
\nn +
\overbrace{ 
\tr \Big( (\wt J_1\wt J_2)^{-1} \d (\wt J_1\wt J_2)  \wedge \wt J_2^{-1} \d \wt J_2 \Big) + \dots }^{\hbox{ contribution at $\wt v$}} \;.
\ee
The terms indicated by dots  are precisely the terms appearing as  the result of the ``zipping'' of the two edges together because they contain only the product $J_1J_2$.
The two terms indicated above cancel out:
$$
\tr \Big((J_1J_2)^{-1} \d (J_1J_2)  \wedge J_2^{-1} \d J_2 
+
 (\wt J_1\wt J_2)^{-1} \d (\wt J_1\wt J_2) \wedge \wt J_2^{-1} \d \wt J_2 \Big) 
$$
$$
=\tr \Big(
 (J_1J_2)^{-1} \d (J_1J_2) \wedge J_2^{-1} \d J_2
 -
  ( J_1J_2) \d ( J_2^{-1} J_1^{-1}) \wedge  \d  J_1 J_1^{-1} \Big) 
$$
$$
=\tr \Big( J_1^{-1} \d J_1 \wedge \d J_2J_2^{-1}+   \d  J_2 J_2^{-1} \wedge J_1^{-1}\d  J_1 \Big) =0
$$
where we have used the cyclicity of the trace  and that $\tr\Big(J^{-1} \d J \wedge J^{-1} \d J \Big)=0$ for any  matrix $J$.
 The computation is almost identical for the case $v=v'$. 
\QED
\begin{lemma}[''Detaching/attaching'' lemma]
\label{edgedetach}
Suppose that two  consecutive edges $e,e'$ at a vertex $v\in {\bf V}(\Sigma)$ (in counterclockwise order) are oriented away and satisfy the relation  $J(e)= J(e')^{-1}$. Let us ``detach'' the edge from the vertex $v$, which then becomes a vertex of valence $n_v-2$ as depicted in Fig. \ref{detach}. Let $(\wt \Sigma, \wt J)$ be the new  pair. Then  $\Omega(\Sigma)=\Omega(\wt \Sigma)$.  Vice versa, we can ``attach'' an edge to a vertex by the inverse procedure preserving the form $\Omega$.
\end{lemma}
{\bf Proof.}
Let $J = J(e)$ and $J(e') = J^{-1}$. For simplicity we assume that $e$ is the first edge in the contribution of the vertex $v$ to the form $\Omega(\Sigma)$. Then this contribution is given by
\be
\tr \Big(  J_1^{-1} \d J_1 \wedge J_1^{-1} \d J_1 
 -
 (J_1J_2)^{-1} \d (J_1J_2)\wedge  J_2^{-1} \d J_2 \Big)  + \dots
\ee
Since $J_1J_2 = J J^{-1} =\1$ the second term vanishes, and the first term vanishes under the trace. The remaining terms give  the contribution of the vertex $v$ without the jump matrices from the first two edges. \QED

\begin{proposition}[Merging of two vertices]\label{edgedeg}
Consider two vertices $u,v\in {\bf V}(\Sigma)$ of valence $\geq 2$ connected by an edge $e$. Denote the valence of
 $u$ by $p+1$ and the valence of $v$ by $q+1$. The jump matrices on the $p$ remaining edges outgoing from $u$ (in counterclockwise order starting from $e$) we denote by $J_1,\dots,J_p$.  We denote by $F_1,\dots,F_q$ the jump matrices on the remaining $q$ edges outgoing from $v$ (in counterclockwise order starting from $e$); due to the condition (\ref{nomono}) at $u$ and $v$ one has
  \be
 J_1\dots J_p\, F_1\dots F_q=\1\;.
 \la{FFGG}\ee
  Denote by $\wt{\Sigma}$ the graph obtained by contracting the edge $e$; under such move  the vertices $u$ and $v$ merge
 forming a vertex  of $\wt{\Sigma}$ of valence $p+q$ which we denote by $w$. Then the forms
 $\Omega(\Sigma)$ and  $\Omega(\wt{\Sigma})$ coincide.
 \end{proposition}
 \begin{figure}[htb]
\begin{center}
\includegraphics[width=0.9\textwidth]{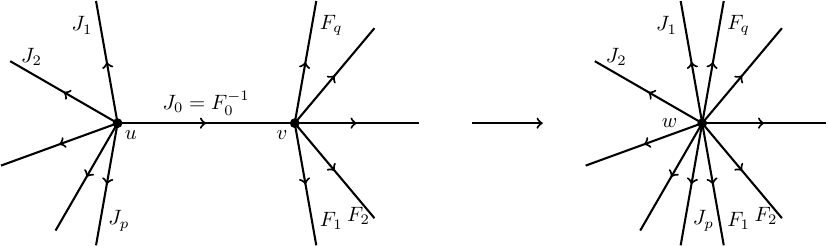}
\end{center}
\caption{The merging of two vertices. }
\la{merging}
\end{figure}

 {\bf Proof.} The contribution of vertices $u$ and $v$ into  $\Omega (\Sigma)$
 is given by
 \be
 &\sum_{l=1}^p  \tr \big(
 (J_1\dots J_\ell)^{-1}\d  (J_1\dots J_\ell)
 \wedge 
 J_\ell^{-1} \d J_\ell
 \big) 
 + \sum_{k=1}^q  \tr \big(
 (F_1\dots F_k)^{-1}\d  (F_1\dots F_k) \wedge F_k^{-1} \d F_k \big)\;.
 \la{uv}
 \ee
 The contribution of the vertex $w$ to $\Omega (\wt \Sigma)$ equals
\be
& \sum_{l=1}^p   \tr \big(
  (J_1\dots J_\ell)^{-1}\d  (J_1\dots J_\ell)\big)\wedge J_\ell^{-1} \d J_\ell
\nn\\
&+  \sum_{k=1}^{q-1}  \tr \big( 
  (J_1\dots J_p F_1\dots F_k)^{-1}\d  (J_1\dots J_p F_1\dots F_k) \wedge F_k^{-1} \d _k F \big)\;.
 \la{w}
 \ee
 The first sums in (\ref{uv}) and (\ref{w}) coincide; taking into account (\ref{FFGG}) one can eliminate all $J_\ell$ to get
\be
 \Omega(\Sigma)-\Omega (\tilde{\Sigma})
 =
 \tr \le( \sum_{k=1}^q  
 (F_1\!\cdots \!F_k)^{-1}\d  (F_1\!\dots\! F_k) \wedge F_k^{-1} \d F_k 
 -
\sum_{k=1}^{q-1} 
\d  (F_{k+1} \!\dots\! F_q) (F_{k+1}\!\dots\! F_q)^{-1} \wedge  F_k^{-1} \d F_k\ri)\;.
 \la{etaeta}
 \ee
 In the first sum the terms containing $F_q$ arise only for $k=q$:
\bea
  \nn \tr &  \le( (F_1\!\cdots\! F_q)^{-1}\d  (F_1\!\cdots\! F_q) \wedge  F_q^{-1} \d F_q\ri)
 =
 \tr\Bigg(   F_q^{-1} \d F_q\wedge F_q^{-1} \d F_q 
\\
&- \sum_{k=1}^{q-1} 
 F_{k+1}\dots F_q F_q^{-1} \d F_q (F_{k+1}\dots F_q)^{-1} \wedge  F_k^{-1} \d F_k\Bigg).
 \la{1sum}
 \eea
 The first term in \eqref{1sum} vanishes due to skew-symmetry of $\wedge$ and the cyclicity of the trace.  In the second sum of (\ref{etaeta}) the terms containing $\d F_q$ are given by 
 $$
  \sum_{k=1}^{q-1} \tr \big( 
  (F_{k+1}\dots F_q) F_q^{-1}\d F_q \wedge  (F_k\dots F_q)^{-1}\d F_k   \big)
  $$
  $$
  =
  -\sum_{k=1}^{q-1}  \tr \big(
   (F_k\dots F_q)^{-1} \d F_k (F_{k+1}\dots F_q) \wedge F_q^{-1} \d F_q \big)
   $$
 which cancels the second term in (\ref{1sum}).
 
 The  terms in  $\Omega(\Sigma)-\Omega(\wt{\Sigma})$,  that do not involve $\d F_q$, are given by the combination
 \bea
 \tr\sum_{k=1}^{q-1}\left(\sum_{\ell=1}^{k-1}
 (F_{\ell+1}\dots F_k)^{-1} F_\ell^{-1} \d F_\ell  \wedge (F_{\ell+1}\dots F_k)F_k^{-1}\d F_k 
 \right.+
 \nn\\
 \left.+
 \sum_{\ell=k+1}^{q-1}
    (F_{k+1}\dots F_\ell) F_\ell^{-1} \d F_\ell \wedge (F_{k+1}\dots F_\ell)^{-1}F_k^{-1}\d F_k\right)
 \eea
  which vanishes due to skew-symmetry in $k$ and $\ell$.
 \QED

\begin{figure}[htb]
\begin{center}
\begin{minipage}{0.69\textwidth}
\begin{center}
\includegraphics[width=0.25\textwidth]{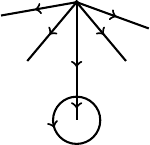}
\\[32pt]
\phantom{0}
\end{center}
\end{minipage} \ \ 
\begin{minipage}{0.50\textwidth}
\includegraphics[width=0.9\textwidth]{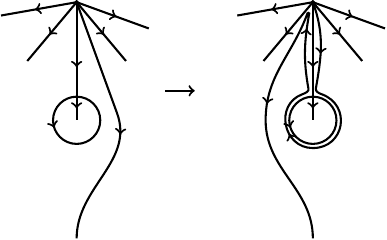}
\end{minipage}
\end{center}
\caption{Top pane: a cherry on the  stem. Bottom pane: cherry movement to the neighbouring face. }
\label{cherry}
\end{figure}

 The graphs of interest in the rest of the paper have an additional structure: the {\it cherry} attached to a vertex $v$ via the {\it stem}. The cherry is constructed as follows: one introduces a one-valent vertex $t$ connected to $v$ by an arc and adds a small counterclockwise loop around $t$ (the  cherry) intersecting transversally the arc. This introduces a four-valent vertex on such arc and split the arc into two edges, the exterior of which we call  {\it stem of the cherry}, see Fig. \ref{cherry}, left pane.

{ The next proposition shows that the cherry 
can be moved from one face of the graph to another without changing the symplectic form.}

\begin{proposition}[Cherry migration]\la{movcher}
The form $\Omega(\Sigma)$ (\ref{dThgen}) remains invariant  if one of the cherries is moved to a neighbouring face.
More precisely, let $e$  be the edge to the right of the cherry,  $J$ be  the jump associated to it (see Fig. \ref{cherry}),  $J_0$ be the jump on the stem  and $C$ the jump on the cherry. Let us   move  the edge $e$ to the left of the cherry  while transforming the jump matrices $(J_0, C)$ to  $(\wt J_0, \wt C)$ where
 $$
 \wt J_0 = J^{-1} J_0 J\;,\hskip0.7cm \wt C = J^{-1} C\;.
 $$
Then the form $\Omega(\wt \Sigma)$ (\ref{dThgen}) coincides with $\Omega(\Sigma)$.
\end{proposition}
\hskip0.5cm

{\bf Proof.}
Using the attaching Lemma  \ref{edgedetach} and the zipping Lemma \ref{zipedge} we can wrap the edge $e_1$ with jump $J_1$ around the cherry from the right to the left, and attach it to the distal vertex of the stem as shown in  Fig.\ref{cherry}. As a result, the jump on the cherry becomes
\be
\tilde{C}=J_1^{-1}C
\la{Ct}
\ee
while the jump $J_0$ on the stem becomes $\wt J_0 = J_1^{-1} J_0 J_1$. 
\QED

As an immediate corollary of Proposition \ref{edgedeg} we get another convenient statement

\begin{figure}[htb]
\begin{center}
\includegraphics[width=0.8\textwidth]{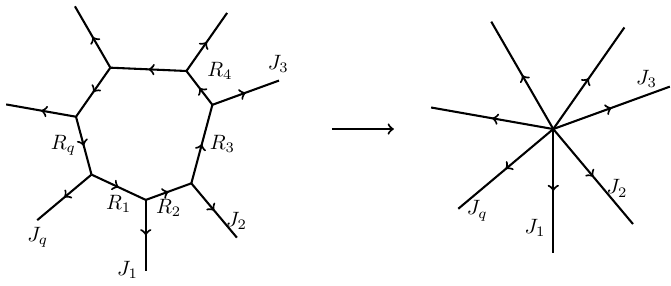}
\end{center}
\caption{Contracting $q$-gonal face to a vertex of valence $q$.}
\label{res_vertex}
\end{figure}

\begin{corollary}[Face contraction]
The form $\Omega(\Sigma)$ remains the same  if one replaces a 
$q$-gonal face by $q$-valent  vertex 
 while  preserving the jump matrices along the $q$ outgoing edges as shown in  Fig.\ref{res_vertex}.

\end{corollary} 

\section{The extended Goldman form as $\Omega(\Sigma_{AM})$}
\label{secAM}

\begin{figure}[htp]
\begin{center}
\includegraphics[width=0.7\textwidth]{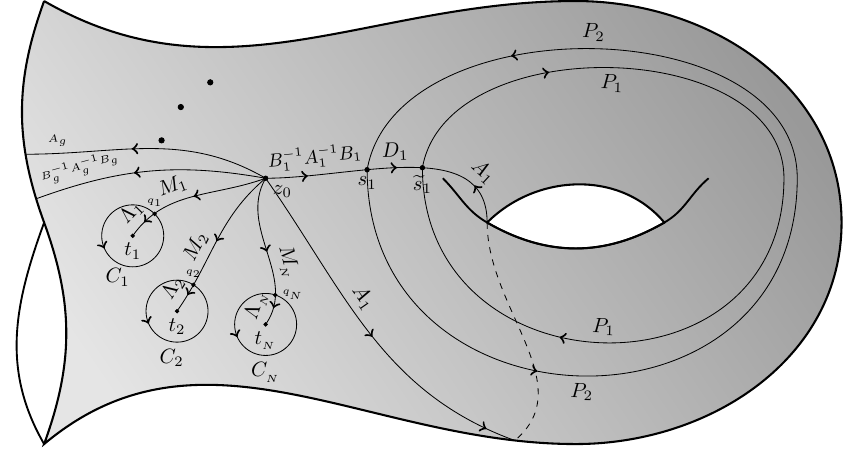}
\vskip1.2cm
\includegraphics[width=0.7\textwidth]{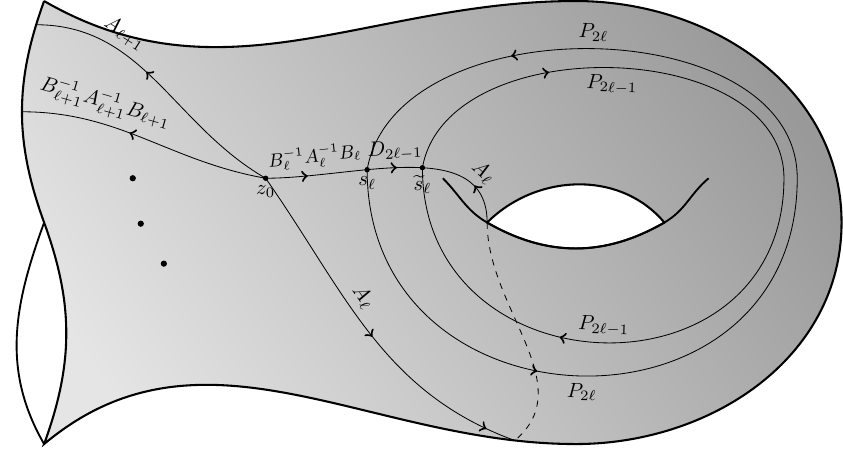}
\end{center}
\caption{The graph $\Sigma_{AM}$ and jump matrices on its edges. On top we show the first handle with jump matrices on the edges of $\Sigma_{AM}$; at the bottom we show  the generic $\ell$th handle for $\ell>1$.}
\label{figAMzip}
\end{figure}

Consider the form $\Omega(\Sigma_{AM})$ with matrices on the edges of $\Sigma_{AM}$ indicated  in Fig. \ref{figAMzip}.

The jump matrices $C_j$ on cherries are matrices diagonalizing the monodromies $M_j$ (\ref{MC}).
The jump  matrices $\Lambda_j$ on segments inside of the cherries are the corresponding diagonal forms of $M_j$ while the jump matrices on the stems are given by $M_j$ themselves. The matrices $A_1$ and $B_1$ are the monodromy matrices (\ref{relfun}) along the canonical generators of the fundamental group. 
The matrices $P_1$ and $D_1$ provide diagonalization of $A_1$ (\ref{diaA}) while the matrix $P_2=B_1^{-1} P_1$ is defined by (\ref{PBP}). The first handle and the corresponding jump matrices are shown in the top of Fig.\ref{figAMzip}.

Similarly, the handle number $\ell$ carries the edges with jump matrices $A_\ell$, $D_{2\ell-1}$, $P_{2\ell-1}$,
$P_{2\ell}$ and $B_\ell^{-1} A_\ell^{-1} B_\ell$ as shown in Fig.\ref{figAMzip}, bottom.

  The notations for the edge  matrices  indicated  in Fig.\ref{figAMzip}   are the same  as the ones used in Section \ref{secinv}. Now we are in a position to formulate the main theorem of this section.
  
\begin{theorem}
\label{thmWAM}
The form ${\mathcal W}$  \eqref{defCh} is related to the  form $\Omega(\Sigma_{AM})$ as follows:
\be
{\mathcal W}= \f{1}{2}\Omega(\Sigma_{AM})\;.
\la{OmW}
\ee
\end{theorem}
{\bf Proof.}
The proof is a direct computation using the general definition \eqref{dThgen} of $\Omega(\Sigma)$. 
Namely, there are $N + 2g +1$ vertices ${\bf V}(\Sigma_{AM})$ (excluding the univalent vertices $t_j$).
The contribution of the vertex  $z_0$ of valence $N+2g$ to $\Omega(\Sigma_{AM})$  \eqref{dThgen}  
 is given by the first term in \eqref{OMAM}.  This is seen by noticing that the  matrices on the corresponding edges are $J_{j} = M_{j}$, $j=1,\dots, N$  and then $J_{N+1} = {A_1}$, $J_{N+2} ={ B_1^{-1}A_1^{-1}B_1}$ and so on. Using the notation $K_{\ell} = J_{1} \cdots J_{\ell}$ (which coincides with the definition \eqref{defK}) the corresponding term in \eqref{dThgen} can be written as  
$$
\sum_{\ell=1}^{n_v}  \tr \big( K_\ell^{-1} \d K_\ell \wedge J_\ell^{-1} \d J_{\ell} \big)\;.
$$
Now observe that $J_\ell = K_{\ell-1}^{-1} K_{\ell}$; thus
$$\d J_{\ell} = -K_{\ell-1}^{-1}\d K_{\ell-1} K_{\ell-1}^{-1}K_\ell  + K_{\ell-1}^{-1} \d K_{\ell}\;.$$
Using this relation we get
$$
\sum_{\ell=1}^{n_v}  \tr  \bigg(
 K_\ell^{-1} \d K_\ell\wedge J_\ell^{-1} \d J_{\ell}  \bigg)
$$
$$
=\sum_{\ell=1}^{n_v}   \tr\bigg( 
 K_\ell^{-1} \d K_\ell  \wedge K_{\ell} ^{-1}K_{\ell-1}\Big( -K_{\ell-1}^{-1}\d K_{\ell-1} K_{\ell-1}^{-1}K_\ell  + K_{\ell-1}^{-1} \d K_{\ell}\Big) \bigg)
$$
$$
  =-\sum_{\ell=1}^{n_v}  \tr\bigg( 
  \d K_\ell K_{\ell} ^{-1} \wedge \d K^{}_{\ell-1} K_{\ell-1}^{-1} 
  \bigg) 
=\sum_{\ell=1}^{n_v}  \tr\bigg( 
  \d K_{\ell-1}^{} K_{\ell-1} ^{-1} \wedge \d K^{}_{\ell} K_{\ell}^{-1}
  \bigg) 
$$
which coincides with  (twice) the first term in \eqref{OMAM}. 

There are now $2g+N$ other contributions to \eqref{dThgen} which arise from the remaining four-valent vertices. They are all of a similar nature;  the four  matrices on the edges attached to these vertices are 
of the type
$$
J_1 = PD^{-1}P^{-1},\ \  J_2 = P,\ \ \ J_3 = D,\  \ \ J_4=P^{-1}
$$
where $D$ is a diagonal matrix and $P\in SL(n)$.  The contribution to the form \eqref{dThgen}  from such a vertex equals 
$$
\tr\big (  
 J_{[1,2]}^{-1} \d J_{[1,2]}\wedge J_2^{-1} \d J_{2}
  +
    J_{[1..3]}^{-1} \d J_{[1..3]}  \wedge J_{3}^{-1} \d J_3 
    \big)
$$
$$
=\tr\le (  
 \Big( D P^{-1} \d (P D^{-1}) \Big) \wedge P^{-1} \d P 
   + 
     P^{-1}\d P \wedge D^{-1} \d D  \ri)
     $$
     $$
  =\tr\le (   
  D P^{-1} \d P D^{-1}\!\! \wedge \! P^{-1} \d P  
  -
   D^{-1} \d D \! \wedge\! P^{-1}  \d P
   +
    P^{-1}  \d P \!\wedge\! D^{-1} \d D  \ri)
    $$
    $$
   = \tr\Big (  DP^{-1}   \d P  \wedge D^{-1}   P^{-1}\d P \Big) 
   +
   2\,\tr \Big(
    P^{-1} \d P  \wedge D^{-1} \d D  \Big)\;.
  $$
From Fig. \ref{figAMzip} we see that the contributions of the vertices $s_{\ell}, \wt s_{\ell}$, $\ell=1,\dots, g$  give the terms contained in the second line of \eqref{OMAM}. The contribution of the vertices $q_j,\ j=1,\dots, N$ gives the second term in the first line of \eqref{OMAM} plus the last term in \eqref{defCh}.  \QED

\begin{remark}\rm
The relationship between our notation and notations of
 \cite{AM} are summarized in the following table:
\begin{center} \begin{tabular}{c|c|c }
Alekseev-Malkin \cite{AM} & Notations of this paper &  range\\
\hline 
$C_j$ & $\Lambda_j$ &   $1\leq j \leq N$\\
$u_j$ & $C_j^{-1}$ & $ 1 \leq j \leq N$\\
$C_{N+\ell}$ & $D_\ell$& $1 \leq \ell \leq 2g$\\
$ u_{N+\ell } $ & $P_{\ell}^{-1}$ & $1 \leq \ell \leq 2g$\\
\hline
\end{tabular}
\end{center}
\end{remark}

\section{The Form  $\Omega(\Sigma_{_{FG}})$}
\la{FGform}

To define the Fock-Goncharov coordinates we  introduce the following auxiliary oriented graphs (see Fig. \ref{figtrian}):
\begin{enumerate}
\item
The oriented graph $\Sigma_0$ with $N$ vertices $v_{1},\dots, v_{N}$ which defines a   triangulation of the  surface; we assume that each vertex $v_j$ lies in a small neighbourhood of the corresponding puncture $t_j$.   Since $\Sigma_0$ is a triangulation there are 
 $2N-4+4g$ faces (triangles) $\{f_k\}_{k=1}^{2N-4+4g}$  and $3N-6+6g$ oriented edges $\{e_k\}_{k=1}^{3N-6+6g}$.

\item
Connect $t_j$ to $v_j$ by an arc and add  a small counterclockwise loop around each $t_k$ (the {\it cherry}) intersecting transversally the arc. This introduces a four-valent vertex on such arc and split the arc into two edges, the exterior of which we called  {\it stem of the cherry} earlier. The cherries are constructed so that  they do not intersect the edges of $\Sigma_0$. The union of $\Sigma_0$, the stems and the cherries is denoted by $\Sigma_1$. This is the black and blue components (as shown in the online version) of the graph in Fig.\ref{figtrian}.

The graph $\Sigma_1$ is determined by $\Sigma_0$ if one chooses  a {\it  ciliation} (following the  terminology of \cite{FR}) 
at each vertex of the graph $\Sigma_0$; the ciliation determines the position of the 
stem of the corresponding cherry.

\item

Choose a point $p_{f_k}, \ \ k=1\dots 2N-4$  inside each triangle $f_k$ of $\Sigma_0$  and connect them by edges $\Ec_{f_k}^{(i)}$, $i=1,2,3$,  oriented towards the point $p_{f_k}$.  We will denote by $\Sigma_{_{FG}}$ the graph resulted by the augmentation of $\Sigma_1$ and these new edges. For definiteness  we will assume that the cherry is always placed between an edge of $\Sigma_0$  on the right and one of the $\Ec_f^{(j)}$'s  on the left, as depicted in Fig.\ref{figtrian}.

\end{enumerate}

\subsection{Fock-Goncharov coordinates}
  \begin{figure}
    \begin{center}
\includegraphics[width=0.7\textwidth]{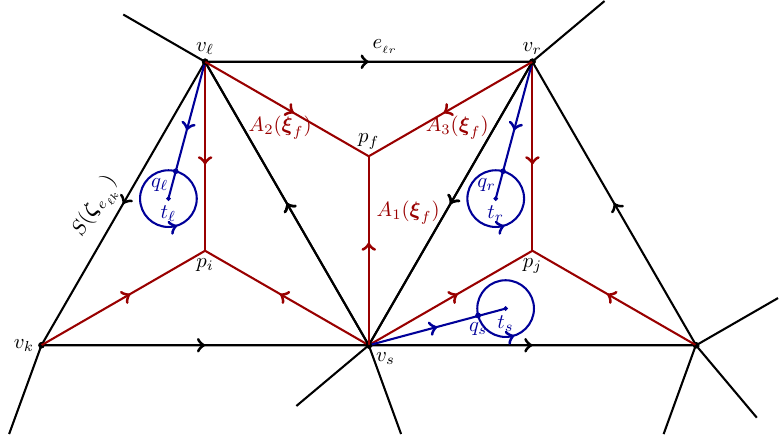}
    \end{center}
    \caption{The support of the jump matrices $J$ is the  graph $\Sigma_{FG}$. The triangulation  $\Sigma_0$  consists of the collection of edges between the vertices $v_j$'s (shown in black in the online version). The graph $\Sigma_1$ is the union of $\Sigma_0$, the ``stems" and ``cherries".  The (shown in red in the online version) edge $\Ec_f^{(\ell)}$ is the edge between $p_i$ and $v_\ell$ inside the face (triangle) $f$.}
    \label{figtrian}
    \end{figure}

Let us denote by $\alpha_i$, $i=1,\dots, n-1$   the simple positive roots of $SL(n)$; by  ${\mathrm h}_i$  we denote the dual roots: 
$$
\alpha_i:= {\rm diag}( 0,\dots, \!\!\!\mathop{1}^{i-pos}\!\!\!,-1,0,\dots),\qquad
{\mathrm h}_i := \le(
\begin{array}{cc}
(n-i)\1_i & 0\\
0& -i\1_{n-i}
\end{array}
\ri)\;,
$$
\be
\tr (\alpha_i {\mathrm h}_k) = n \delta_{ik}\;.
\label{rootprods}
\ee
For  any matrix $M$ we define $M^\star:= P M P$ where $P$ is the ``long permutation" in the Weyl group, 
$$P_{ab}= \delta_{a,n+1-b}\; \ \ \ {\rm i.e.} \ \ \
P= \le[\begin{array}{ccccc}
0&0 &\cdots& 0& 1\\
0&0&\cdots&1&0\\
\vdots & &\hbox{\reflectbox{$\ddots$}} && \vdots\\
0&1&\cdots&0&0\\
 1&0&\cdots &0&0
\end{array}\ri]\;.$$
In particular,
$\alpha_i^\star = -\alpha_{n-i}$ and $ {\mathrm h}_i^\star = -{\mathrm h}_{n-i}\;.$

The main ingredient of the formulas below is the matrix  $\CM$  given by 
\be
\CM_{jk}=\tr ({\mathrm h}_j {\mathrm h}_k)=n^2\left({\rm min}(j,k)-\f{jk}{n}\right)
\la{defCM}
\ee
which equals to $n^2 A_{n-1}^{-1}$ with $A_{n-1}$ being the Cartan matrix of $SL(n)$.

The full set of coordinates on $\Mcal$ consists of three groups: the coordinates assigned to vertices of the
graph $\Sigma_0$, to its edges and faces. Below we describe these three groups separately and use them to parametrize the jump matrices on the edges  of the graph $\Sigma_{FG}$.

\noindent {\bf Edge coordinates and  matrices.}
To each edge $e \in {\bf E} (\Sigma_0)$ we associate $n-1$ complex non-vanishing  variables 
\be
{\bs z} = \bs z_e =  ( z_1, \dots, z_{n-1} )
\ee
and introduce their logarithmic counterparts: 
\be
{\bs \zeta}={\bs \zeta}_e = (\zeta_{1},\dots, \zeta_{n-1})\in \C^{n-1}\;, \qquad \zeta_j=\f{1}{n}\log z_j^n\;.
\la{vare}\ee 
The consideration in this paper is  local. Therefore, our convention about  all logarithms here and below is to
 use the principal branch of the logarithm with the branch cut along $\R_-$. 

The matrix on the oriented edge $e\in {\bf E}(\Sigma_0)$  is given by 
\be
S({\bf z})  &= {\bf z}^{-\boldsymbol h}P \sigma:=  \prod_{j=1}^{n-1} z_j^{-{\mathrm h}_j} P \sigma =
\nn { \prod_{\ell=1}^{n-1} z_\ell^{\ell}} \le(
\begin{array}{ccccc}
0 & \dots & & &  (-1)^{n-1} \prod_{j=1}^{n-1}z_{j}^{-n}\\
 & & & \hbox{\reflectbox{$\ddots$}} & 0\\
\vdots &&&\\
0 &  { -z_{n-2}^{-n}}{z_{n-1}^{-n}} &0 & \dots\\
1 & 0 \dots
\end{array}
\ri)
\ee
where 
$$\sigma ={\rm diag} (1,-1,1,-1,\dots)$$
 is the signature matrix { and the notation ${\bf z}^{\bf h}$ stands for
 {\be
 {\bf z}^{\bf h} = z_1^{\rm h_1}\dots z_{n-1}^{\rm h_{n-1}}
 \ee }
 with ${\mathrm h}_j$ being the simple coroots of $SL(n)$ \eqref{rootprods}.
 }
 For the inverse matrix we have
$$S^{-1}({\bf z}) =\s P {\bf z}^{\bf h}  = (-1)^{n-1} {\bf z}^{\bf h^\star} P \s \;.$$ 
Since 
$${\mathrm h}_i^\star = P {\mathrm h}_i P=- {\mathrm h}_{n-i}\;,$$
the sets of variables (\ref{vare}) corresponding to an oriented edge $e$ of $\Sigma_0$ and the opposite edge $-e$ 
are related as follows: 
\be
\label{edgereverse}
 {\bf z}_{-e}  :=(-1)^{n-1} (z_{e,n-1}, \dots, z_{e,1})\;,\hskip0.8cm
{\bs \zeta}_{-e} = (\zeta_{e,n-1} ,\dots, \zeta_{e,1}) \;.
\ee
Notice that in the $SL(2)$ case $z_{-e}=-z_e$ while $\zeta_{-e}=\zeta_e$.

\noindent {\bf Face coordinates and  matrices on $\Ec_{f}^{(i)}$.}
To each triangle $f\in {\bf F}(\Sigma_0)$  we associate 
$\frac {(n-1)(n-2)}2$   variables ${\bs x}_f = \{x_{f;\,abc}:\ \ a, b, c\in \N,\ \ \ a + b + c= n\}$  and their {logarithmic counterparts }
$$\xi_{f;\,abc}=\log({x_{f;\,abc}})$$ 
 as follows.
The variables $x_{f;\,abc}$ define the  matrices $A_{i}({\bf x}_f)$ on three edges $\{\Ec_{f}^{(i)}\}_{i=1}^3$,  which connect a 
chosen point $p_f$ in each face $f$ of the graph $\Sigma_0$ with its three vertices (these edges are   shown in red (in the online version) in Fig. \ref{figtrian}). The enumeration of vertices $v_1$, $v_2$ and $v_3$  is chosen arbitrarily
for each face $f$.

Namely, for  a given vertex $v$ and the face $f$ of $\Sigma_0$ such that $v\in \pa f$ we define the index $f(v) \in \{1,2,3\}$ depending on the enumeration that we have chosen for the three edges $\{\Ec_{f}^{(i)}\}$ lying in the face $f$. For example  for the face $f $ containing point $p_f$  in Fig.  \ref{figtrian} we define  $f(v_\ell)=2$, $f(v_s)=1$ and $f(v_r) = 3$.   

Let $E_{ik}$ be the elementary matrix and define
$$
F_i  = \1 + E_{i+1,i}\; ,
$$
$$
H_i(x) = x^{-{\mathrm h}_i}={\rm diag}(\overbrace{x^{n-i},\dots, x^{n-i}}^{\hbox{$i$ times}}, x^{-i},\dots x^{-i})  \;,
$$
$$  i=1,\dots, n-1\;;$$
$$
N_k= \le(\prod_{k\leq  i \leq n-2} 
H_{i+1}(x_{n-i-1, i-k+1, k})
F_i\ri) F_{n-1}\; .
$$
Then the matrix $A_1$ is defined as follows \cite{FG} 
\be
A_1({\bs x}) =\sigma\;   \le(\prod_{k=n-1}^1 N_k\ri)\; P\;.
\la{defA1}
\ee
The matrices $A_2$ and $A_3$ are obtained from $A_1$ by cyclically permuting the indices of the variables: 
\be
A_2({\bs x}) = A_1(\{x_{bca}\})\;,\hskip0.7cm
A_3({\bs x}) = A_1(\{x_{cab}\})\;.
\la{defA23}
\ee
The important  property of  the matrices $A_i$ is the equality 
\be
\la{A123}
A_1 A_2 A_3
=\1\; .
\ee
The equation (\ref{A123}) guarantees the triviality of total monodromy around the point $p_f$ on each triangle $f\in {\bf F}(\Sigma_0)$.

Let us now introduce the following diagonal matrices  (the superscript $^D$ indicates the diagonal part of a matrix)
\be
\label{diagA}
{\bf x}^{-{\bf h}_i}=\Big(P \sigma A_i({\bs x})\Big)^D \;  ,\qquad i=1,2,3\;.
\ee
These matrices can be expressed as follows in terms of variables $x_{abc}$:
\bea
\label{facemat}
 {\bf x}^{\bf h_1}= \prod_{a + b + c =n} x_{abc}^{{\mathrm h}_a}\;,\hskip0.7cm
 {\bf x}^{\bf h_2}=
\prod_{a + b + c =n} x_{abc}^{{\mathrm h}_b} \;,\hskip0.7cm
 {\bf x}^{\bf h_3}=
\prod_{a + b + c =n} x_{abc}^{{\mathrm h}_c}  \;.
\eea
\vskip0.5cm

 \begin{example}\rm
 In the first three non-trivial cases the matrices $A_i$ have the following forms:
\begin{enumerate}
\item[$SL(2)$:]
 there are no  face variables and all matrices $A_i=A$ are given by
\be
A=\left( \begin{array}{cc} 0&1\\ -1&-1
\end{array} \right)\;.
\ee
\item[$SL(3)$:]
 there is one  parameter $\xi=\xi_{111}$ for each face. The  matrices $A_1,A_2$ and $A_3$ coincide in this case, too; they are  given by 
\be
A(\xi) =  x  \left( \begin {array}{ccc} 0&0&1\\ [5pt]0&-1&-1
\\ [5pt]{x}^{-3}&{x}^{-3}+1&1\end {array} \right)\; \ ;\ \ \ x = {\rm e}^\xi\;.
\ee
\item[$SL(4)$:]
 the three matrices $A_1,A_2,A_3$  are different   and  $A_1$ is given by 
 \be
\la{ASL4}
\small
A_1({\bs \xi}) = - { x_{211}^2 x^{}_{121} x^{}_{112}} 
\left( \begin {array}{cccr} 
0&0&0& -1\\ 
\noalign{}0&0&1&1
\\ \noalign{} 0&-x_{211}^{-4}& - x_{211}^{-4}-1&-1
\\ \noalign{}
x_{112}^{-4}x_{211}^{-4}x_{121}^{-4} & x_{211
}^{-4} \left( x_{112}^{-4}x_{121}^{-4}+x_{112}^{-4}+1 \right) &
1+ 
 \left( x_{112}^{-4}+1 \right) x_{211}^{-4}&1\end {array} \right)
\ee
where $x_{ijk} = {\rm e}^{\xi_{ijk}}$.
\end{enumerate}
\end{example}
\noindent {\bf Matrices on stems.}
For each vertex $v$ of $\Sigma_0$ of valence $n_v$ the jump  matrix on the stem of the cherry connected to a vertex $v\in {\bf V}(\Sigma_0)$ is defined from the triviality of  total monodromy around $v$ \eqref{nomono} and is given by 
\be
M_v^{0} = \le(\prod_{i=1}^{n_v} A_{f_i}  S_{e_i} \ri)^{-1} 
\la{Mv0}
\ee
where $f_1,\dots, f_{n_v}$ and $e_1,\dots e_{n_v}$ are the faces/edges ordered counterclockwise starting from the stem of the cherry, with the  edges oriented away from the vertex (using if necessary the formula \eqref{edgereverse}).   Since each product $A_{f_i}  S_{e_i}$ is a lower triangular matrix, the matrices $M_v^0$ are also lower--triangular. The diagonal parts of $M^0_v$ will be  denoted by $\Lambda_v$ and parametrized as shown below
\be
\Lambda_v
= {\rm diag} \le(m_{v;1} , \frac {m_{v;2}}{m_{v;1}}, \dots, \frac {m_{v;n-1}}{m_{v;n-2}}, \frac 1 { m_{v;n-1}}\ri)\;.
\label{moneig}
\ee  
Notice that the  matrix (\ref{moneig}) can be written as  ${\bf m}_v^{\boldsymbol \a}$ where 
${\bf m}_v={\rm diag} \le(m_{v;1},\dots,m_{v;1}\ri)$.

In order to express $\Lambda_v$ in terms of $z$- and $x$ -coordinates , we enumerate the faces and edges incident at the vertex $v$ by $f_1,\dots, f_{n_v}$ and $e_1,\dots, e_{n_v}$, respectively. 
We also assume without loss of generality  that the arc $\mathcal E_{f_j}^{(1)}$ is the one connected to the vertex $v$ for all $j=1,\dots, n_v$. Then, using \eqref{diagA}  we obtain the formula
\be
\Lambda_v = P  \left(\prod_{f\perp v} {\bf x}_f^{{\bf h_1}}\right)\left(  \prod_{e\perp v}  z_{e}^{{\bf h}} \right) P  \;.
\la{Lambdaxz}
\ee

Let us now define also the variables
\be
\label{muvj}
\mu_{v;n-\ell} ={ \frac 1 n}\le(\sum_{f\perp v} \sum_{a+b+c=n\atop a,b,c\geq 1} \xi_{f;abc}\, \CM_{a\ell}
+ \sum_{e\perp v} \sum_{j=1}^{n-1} \zeta_{e;j}\, \CM_{j\ell}\ri)
\ee
where the matrix $\CM$ equals to $n^2$ times the inverse Cartan matrix (see (\ref{defCM})).

Up to an $n$th roots of unity, the exponents of the variables $\mu$ give variables $m$ from (\ref{moneig}), namely,
$$
m_{v;\ell}^n =  {\rm e}^{n\mu_{v;\ell}}\;.
$$

\noindent {\bf Vertex coordinates and  matrices on cherries.}
To each vertex $v$ of the graph $\Sigma_0$ we associate a set of $n-1$ {\it toric} or {\it  vertex} coordinates
$r_{v;i} \in \C^\times $, $i=1,\dots,n-1$ as follows.
Since the matrix $M^0_v$ is  lower-triangular it can be diagonalized by a lower-triangular matrix $C_v^0$
such that all diagonal entries of $C_v^0$ are equal to  1:
\be
M^0_v=C_v^0\Lambda_v (C_v^0)^{-1}\;.
\la{diaM0}
\ee
Any other lower-triangular matrix $C_v$ diagonalizing $M^0_v$ can be written as
\be
C_v= C_v^0 \verB_v
\la{CvCv0}\ee
where the matrix $\verB_v$  equals to the diagonal part of $C_v$, $\verB_v=(C_v)^D$. The matrix $\verB_v$
is parametrized by $n-1$ 
variables $r_1, \dots, r_{n-1}$ as follows:
$$
\verB = \prod_{i=1}^{n-1} \ver_{i}^{ {{\mathrm h}_i}}  = {\bf \ver} ^{\bf h}
$$
\be
=\left(\prod_{i=1}^{n-1} \ver_i^i\right)^{-1} \!\!\!\!\!\! {\rm diag} \Big(\prod_{i=1}^{n-1} \ver_i^n, \,\prod_{i=2}^{n-1} \ver_i^n\;,\, \dots,\ver_{n-2}^n\ver_{n-1}^n,\,  \ver_{n-1}^n,\, 1\Big)
\la{parR}
\ee
where the set of variables $\{\ver_j\}$ depends on the vertex but we have omitted the corresponding subscript here for readability. 
 The matrix on the cherry attached to the vertex $v$ via stem is defined to be 
 $ J_v=C_v  $.

We define also logarithmic partners of $r_j$'s  via
\be
\rho_{v;j}=\f{1}{n}\log r_{v;j}^n\;;
\la{rhor}
\ee
again it is convenient to assign the same $\rho$-variable to  $r$-variables which differ by the $n$th root of unity.

\subsection{Computation of the form $\Omega(\Sigma_{_{FG}} )$}
\label{secsymp}
The goal of this section is 
to express the  symplectic form $\Omega(\Sigma_{_{FG}})$  in the  coordinates $\{{\bs \xi}, {\bs \zeta}, {\bs \rho}\}$ introduced in the previous section.
The form $\Omega(\Sigma_{_{FG}})$ equals to the  sum of several contributions from  the vertices $v\in {\bf V}(\Sigma_0)$ of the  triangulation $\Sigma_0$ (shown in black in Fig.  \ref{figtrian})  and the vertices $p_f$ at the centers of the triangles  $f\in {\bf F}(\Sigma_0)$.  Contributions of these vertices can be understood also as contributions of the faces of $\Sigma_0$. We start from the following proposition which will be used to compute  the contributions of vertices $p_f$.

\begin{proposition}
\label{propomegaf}
Let matrices $A_{1,2,3}$ be expressed via coordinates $x_{ijk} = {\rm e}^{\xi_{ijk}}$, associated to a face $f$ of the graph $\Sigma_0$,  by  (\ref{defA1}), (\ref{defA23}).
Then the contribution to $\Omega(\Sigma_{_{FG}})$ coming from the vertex at the center of the triangle $f\in {\bf  F}(\Sigma_0)$ is the form  
\be
\omega_f=\tr\bigg(  \d A_2 A_2^{-1}  \wedge A_1^{-1}\d A_1 \bigg).
\la{omfdef}
\ee
It
can be equivalently represented as follows
\bea
\label{omegaf}
\omega_f
=\sum_{i + j + k = n\atop 
i' + j' + k' = n}
F_{ijk; i'j'k'} \,\, \d \xi_{f; ijk}    \wedge \d \xi_{f; i'j'k'}
\eea
where $F_{ijk; i'j'k'}$ are the following constants
\be
&F_{ijk;i'j'l'}=    
(\CM_{i,n-j'} - \CM_{i',n-j}    ) H(\Delta i \Delta j)
  \la{coefface}
  +  (\CM_{j,n-k'} - \CM_{j',n-k}   ) H(\Delta j \Delta k)+
 (\CM_{k,n-i'} -\CM_{k',n-i}   ) H(\Delta k \Delta i)
\ee
where 
$$\Delta i= i'-i\;, \hskip0.7cm
\Delta j= j'-j\;,  \hskip0.7cm \Delta k= k'-k\;,
$$
$\CM$ is given by (\ref{defCM}) and $H(x)$ is the Heaviside function:
\be
H(x) = \le\{
\ba{cc}
1  & x>0\\
\frac 1 2  &  x=0\\
0 & x<0
\ea\ri.\;.
\ee
\end{proposition}
\begin{remark}\rm
 The expression (\ref{coefface}) can be equivalently written  as follows
$$
\frac 1 n F_{ijk; i'j'k'}= ( j'k - k'  j ) H( \Delta j \Delta k )  
$$
$$
+( k' i -i'  k ) H( \Delta i \Delta k )+(  i'  j -j'  i) H( \Delta i \Delta j )\; .
$$
Note that due to the condition $\Delta i + \Delta j + \Delta k =0$, there is  always  a pair of the variables $i,j,k$ (possibly two pairs) such that $\Delta i \Delta j\geq 0$. If the inequality is strict there is exactly one pair. If one of the $\Delta$'s is zero, then there are two pairs with this property.
\er

\noindent {\bf Proof.} 
The three jump matrices on the edges meeting at $p_f$, with the edges oriented outwards are $J_1 = A_1^{-1}$, $J_2 = A_3^{-1}$, $J_3= A_2^{-1}$. Then the vertex contribution, keeping in mind that $J_1J_2J_3 =\1$ (which follows from \eqref{A123}), boils down to a single term that can be written in any of the  three equivalent forms:
$$
\omega_f = \tr\bigg(  \d A_2 A_2^{-1} \wedge  A_1^{-1}\d A_1\bigg)
=\tr\bigg(  \d A_3 A_3^{-1} \wedge A_2^{-1}\d A_2\bigg)
$$
$$
=  \tr\bigg(
  \d A_1 A_1^{-1} \wedge A_3^{-1}\d A_3  \bigg)\;.
$$
For convenience  we use the last expression in the following computations.
Let us now compute  $\omega_f( \pa_{\xi_{ijk}}, \pa_{\xi_{i'j'k'}})$.
The proof of the following lemma is straightforward:
\bl
The matrix $\pa_{_{\xi_{ijk}}} A_1 A_1^{-1} $ is lower triangular; the nontrivial  entries in the lower-triangular part are confined in the region indicated in  Fig. \ref{figshape}.
Similarly, $A_1^{-1}\pa_{\xi_{ijk}} A_1  $ is an upper triangular matrix of the indicated shape.
For $A_2, A_3$ the same statements hold with $(i,j,k)$ replaced by $(j,k,i)$ and  $(k,i,j)$  respectively.
\begin{figure}
\begin{center}
\includegraphics[width=0.8\textwidth]{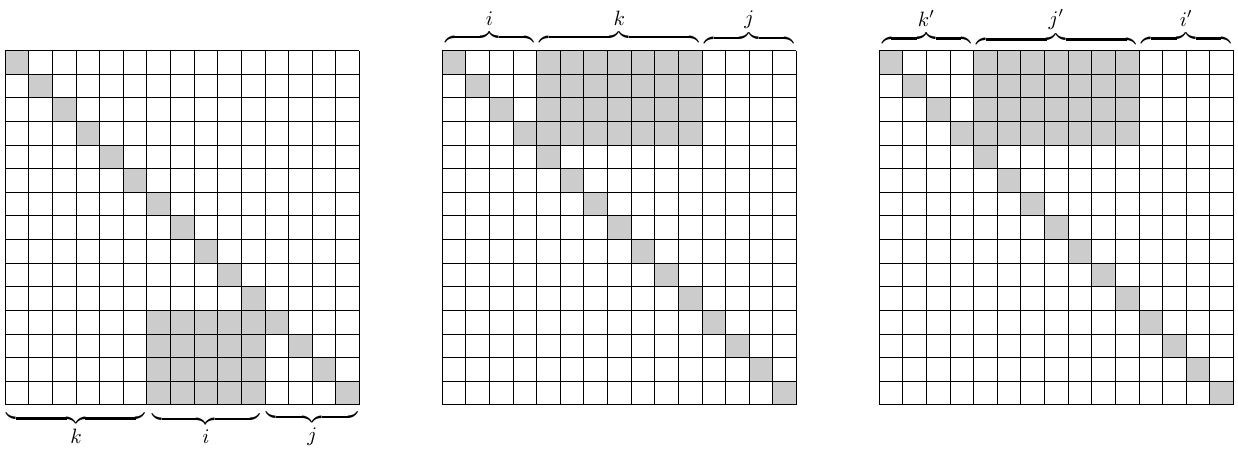}
\end{center}
\caption{The shapes of matrices $\pa_{{\xi_{ijk}}} A_1 A_1^{-1} $,  $ A_1^{-1}\pa_{_{\xi_{ijk}}} A_1$ and $  A_3^{-1} \pa_{\xi_{i'j'k'}} A_3 $, respectively. Non-vanishing entries are confined by the shaded regions.}
\label{figshape} 
\end{figure}
\el

Consider  now $ \tr\bigg( \pa_{\xi_{ijk}}  A_1 A_1^{-1}   A_3^{-1}\pa_{ \xi_{i'j'k'}} A_3\bigg)$; the shapes of the two matrices involved in this expression are  shown in Fig.  \ref{figshape}.
The entries of the blocks outside of the diagonal contribute to the diagonal entries of the product only if 
\be
k<k'\ , \ \ i' <  i \ \ \Rightarrow \ \  \Delta i \Delta k<0;
\ee
this condition is invariant under the exchange $i\leftrightarrow i', j\leftrightarrow j', k\leftrightarrow k'$. 

Suppose now that $\Delta i \Delta k\geq 0$ so that only the diagonal entries  of $ \d A_1 A_1^{-1}$ and $ A_3^{-1} \d A_3$ are involved in the trace. These entries are given by 
$$
(\d A_1A_1^{-1})^D    = \sum_{j=1}^{n-2} \sum_{i=j}^{n-2} \d \log H_{i+1}(x_{n-i-1, i-j+1, j})
=\sum_{a+b+c=n\atop a',b',c'\geq 1} {\mathrm h}_{n-a} \d \xi_{abc}\;,
$$
$$
(A_3^{-1}\d A_3)^D    = 
\sum_{j'=1}^{n-2} \sum_{i'=j'}^{n-2} \d \log H_{n-i'-1}(x_{i'-j'+1, j' ,n-i'-1})
$$
$$
= - \sum_{a'+b'+c'=n\atop a',b',c'\geq 1} {\mathrm h}_{c'} \d \xi_{a'b'c'}\;.
$$
In this proof, the notation $\1_s$ is used for  the diagonal matrix of size $n\times n$ with the identity of size $s$ in the top left block and other entries equal to $0$. The notation $\widetilde \1_s = J\1_s J$ similarly denotes the $n\times n$ diagonal matrix with the identity of size $s$ in the bottom-right block. 

Consider the coefficient in front of $\d \xi_{_{ijk}}\wedge \d \xi_{_{i'j'k'}}$.  This coefficient equals to  the difference of the term  $\tr \big(\pa_{\xi_{ijk}}A_1 A_1^{-1} A_3^{-1}\pa_{\xi_{i'j'k'}} A_3\big)$ and the term 
where  the primed variable are exchanged with the non-primed. The first term is given by 
\bea
-\tr \big(\pa_{_{\xi_{ijk}}}A_1 A_1^{-1} A_3^{-1}\pa_{_{\xi_{i'j'k'}}} A_3\big) =\tr \big( {\mathrm h}_{n-i} {\mathrm h}_{k'}\big)
\la{trAA}.
\eea
Since we are considering  the case $\Delta i \Delta k\geq 0$, we can assume without loss of generality (up to swapping the role of primed and non-primed variables) that $\Delta i,\, \Delta k\geq 0$.
. Then one verifies that the above expression reduces to $ -n \,i\,\Delta k'$. 
Antisymmetrization  gives $n(ik'-i'k)$ which leads to (\ref{coefface}). \QED

 Below we write explicitly the form  (\ref{trAA}) for small $n$.
\begin{example}\rm
For $SL(2)$ and $SL(3)$ the form $\omega_f$ vanishes. 
For $SL(4)$ we get
\be
\frac {\omega_f}{4} =  \d \xi_{211} \wedge  \d \xi_{121}   + \d \xi_{112} \wedge \d \xi_{211}   + 
\d \xi_{121}\wedge \d \xi_{112}\;.
\ee
For
  $SL(5)$ we have 
\bea
\frac {\omega_f}{5}& = 
 \d \xi_{311}\wedge \d \xi_{221}
+ 2 \d \xi_{311} \wedge  \d \xi_{131}
+  \d \xi_{212} \wedge \d \xi_{311}
+2 \d \xi_{113} \wedge \d \xi_{311}
\nn  \\ \nn
&+   \d \xi_{221} \wedge \d \xi_{131} +2\d \xi_{221} \wedge \d \xi_{212}
+2 \d \xi_{122} \wedge \d \xi_{221}
+ \d \xi_{131}  \wedge \d \xi_{122}
\\ \nn
&+2\d \xi_{131} \wedge  \d \xi_{113}
+2\d \xi_{212} \wedge \d \xi_{122}
+ \d \xi_{113} \wedge  \d \xi_{212}
+ \d \xi_{122} \wedge \d \xi_{113}\;.
\eea
For $SL(6)$  the matrix of coefficients is
\bea
\small\nn
\frac{\omega_f}{6} \mapsto \left( \begin {array}{ccccccccccc} &\xi_{411}&\xi_{321}&\xi_{231}&\xi_{141}&
\xi_{312}&\xi_{222}&\xi_{132}&\xi_{213}&\xi_{123}&\xi_{114}
\\[5pt]\xi_{411}  &0&1&2&3&-1&0&1&-2&-1&-3
\\[5pt]\xi_{321}&-1&0&1&2&3 &-2&-1&1&-4&-1
\\[5pt]\xi_{231}&-2&-1&0&1&1&2&-3&4&-1&1
\\[5pt]\xi_{141}&-3&-2&-1&0&-1&0&1&1&2&3
\\[5pt]\xi_{312}&1&-3&-1&1&0&2&4&-1&1&-2
\\[5pt]\xi_{222}&0&2&-2&0&-2&0&2&2&-2&0
\\[5pt]\xi_{132}&-1&1&3&-1&-4&-2&0&-1&1&2
\\[5pt]\xi_{213}&2&-1&-4&-1&1&-2&1&0&3&-1
\\[5pt]\xi_{123}&1&4&1&-2&-1&2&-1&-3&0&1
\\[5pt]\xi_{114}&3&1&-1&-3&2&0&-2&1&-1&0\end {array}
 \right).
\eea
\end{example}
The following  theorem is the main result of this section.
\begin{theorem}
\label{thmtoric}
The symplectic form $\Omega(\Sigma_{_{FG}})$ is expressed by the  following formula
 \bea\la{ototal}
\Omega(\Sigma_{_{FG}}) =  &  \sum_{v\in  {\bf V}(\Sigma_0)} 
\omega_v+ \sum_{f\in  {\bf F}(\Sigma_0)} \omega_f + 2n\!\!\!\! \sum_{v\in {\bf V} (\Sigma_0)} \sum_{i=1}^{n-1} \ {   \d \rho_{v;i} \wedge \d \mu_{v;i} }
\eea
where  $\mu_{v;j}$'s are  defined in \eqref{muvj}. 

The form $\omega_v$ in (\ref{ototal})  is defined as follows:   for each vertex $v\in V(\Sigma_0)$ of valence $n_v$ let $ \{e_1,\dots e_{n_v}\}$ be the incident edges ordered counterclockwise starting from the one on the left of the  stem (and oriented away  from $v$ according to our definition of incidence). Let $\{f_1,\dots, f_{n_v}\}\in F(T)$ be the faces incident to $v$ and counted in counterclockwise order  from the one containing the cherry.  We denote the ordering by $\prec$. Then 
\be \omega_v = &
 \sum_{e'\prec e \perp v} \sum_{i,j=1}^{n-1}   \CM_{ij} \d {\zeta}_{e'; i} \wedge    \d  {\zeta}_{e;j}
+
\sum_{f\prec e\perp v} \sum_{a+b+c =n} \sum_{\ell=1}^{n-1} \CM_{f(v),\ell} \d \xi_{f;abc}   \wedge \d \zeta_{e;\ell} 
\nn\\
&+
\sum_{e\prec f\perp v} \sum_{a+b+c =n}\sum_{\ell=1}^{n-1}  \CM_{f(v),\ell}  \d \zeta_{e;\ell}  \wedge \d \xi_{f;abc} 
+
\sum_{ f' \prec  f\perp v }  \sum_{a+b+c=n\atop a'\!\!+b'\!\!+c'\!=n}\CM_{f'(v),f(v)}  \d {\xi}_{f'; a'b'c'}  \wedge    \d  {\xi}_{f;abc}
\la{omegav}
\ee
where $\CM$ is defined by (\ref{defCM}), the subscript ${f(v)}$ indicates the index $a, b$ or $c$ depending on the value $f(v) \in \{1,2,3\}$, respectively.
The form $\omega_f$ for a face $f$ is given by (\ref{omegaf}).
\end{theorem}

\noindent {\bf Proof.}  
We are going to evaluate all contributions of expression (\ref{dThgen}) in terms of coordinates $\{{\bs \xi}, {\bs \zeta}, {\bs\rho}\}$.
Let us  start from the term  $\omega_v$. 
The contribution  of the vertex $v$ of the graph $\Sigma$   is given by 
\be
\label{locom}
\omega_v =  \sum_{\ell=1}^{n_v-1} \tr \bigg(  J_{[1:\ell]}^{-1} \d J_{[1:\ell]}  \wedge J_\ell^{-1} \d J_\ell\bigg)
\ee
where $J_1,\dots, J_{n_v}$  are the jump matrices of the edge oriented away from $v$ and labeled in counterclockwise order.  Our convention is that the stem of the cherry is followed by one of the $\Ec$ edges (with jump matrix one of $A_{1,2,3}$  so that there are an even number $2n_v$ of edges (except the stem) and the sequence  of the matrices is $A_{f_1(v)} , S_{e_1}, A_{f_2(v)} , S_{e_2}, etc.$, see Fig. \ref{sin_vertex}.

\begin{figure}
\begin{center}
\begin{tikzpicture}[scale=1.6]
   \foreach \i/\J in  { 0/{$J_5=A_{j_5}^{f_5}$}, 1/{$ $}, 2/{$J_{2q-1}=A_{j_{2q-1}}^{f_{2q-1}}$}, 3/{$J_1=A_{j_1}^{f_1}$}, 4/{$J_3=A_{j_3}^{f_3}$} } 
        \draw [very thick,red!60!black,postaction={decorate,decoration={markings,mark=at position 0.5 with {\arrow[line width=1.5pt]{>}}}}] (0,0) to node[pos=0.8, sloped,below]{\J} (72*\i:2);
    
   \foreach  \i/\J in  { 0/{$ $}, 1/{$ $}, 2/{$J_{2q}=S_{e_q}$}, 3/{$J_2=S_{e_1}$},4/{$J_4=S_{e_{_2}}$}}  {
        \draw [very thick,postaction={decorate,decoration={markings,mark=at position 0.5 with {\arrow[line width=1.5pt]{>}}}}] (0,0) to node[pos=0.8, sloped, above] {\J} (36+72*\i:2);
    }
    
 \draw [very thick ,postaction={decorate,decoration={markings,mark=at position 0.5 with {\arrow[line width=1.5pt]{>}}}}] (0,0) to node[pos=0.64, sloped, above] {$J_0$} (-72*2 - 18:2) coordinate (c0);
 \draw [  very thick,postaction={decorate,decoration={markings,mark=at position 0.5 with {\arrow[line width=1.5pt]{>}}}}] (c0)  circle  [radius=0.4];
 \node at ($(c0)+ (180:0.66)$) {$C$};
 \draw [ fill] (c0) circle[radius=0.02] node [below]{$t_k$} ;
 \draw [ fill] ($(c0) + (19:0.4)$) circle[radius=0.02];
 \node at (18:0.37) {$v$};
\end{tikzpicture}

\end{center}
\caption{Contribution of  vertex $v=v_k$.  Here the matrices $A_j({\bs\xi_f})$ are denoted simply by $A_j^f$ for brevity. 
}
\label{sin_vertex}
\end{figure}
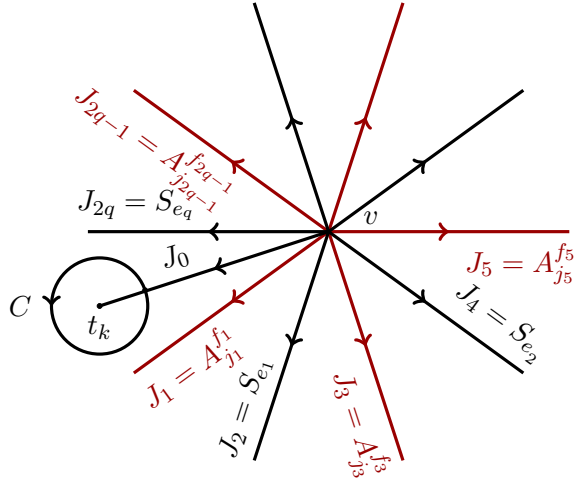

Given the shapes of the face  matrices $A_{1,2,3}$ and the edge matrices  $S_e$, each  addendum in \eqref{locom} is the trace of the wedge product of two lower triangular matrices  (for even $\ell$) or two  upper triangular matrices (for odd $\ell$), and hence only the diagonal entries give non-vanishing contributions.
Since the shape of the matrices $A_{1,2,3}$ is $A = L P = P  U$ with $L$ being a lower-triangular and $U = P  L P $ an upper-triangular matrices, and $S =  {\bf z}^{-\bf h} P  \sigma$, the contribution of the vertex $v$ is given by 
$$
 \omega_v=
\tr   \le( 
\sum_{j=1}^{n_v}  
 \d \log\le(\prod_{f\prec e_j}  {\bf x}_f^{-\bf h_{f(v)}} 
 \prod_{e\prec e_j} {\bf z}_{e}^{-\bf h} \ri ) \wedge  \d \log {\bf z}_{e_j}^{-\bf h} 
 \ri.
 $$
 $$
 + \le. \sum_{j=1}^{n_v} 
   \d \log\le(\prod_{f\prec f_j}  {\bf x}_f^{-\bf h_{f(v)}} \prod_{e\prec f_j} {\bf z}_{e}^{-\bf h} \ri ) \wedge \d \log {\bf x}_{f_j}^{-\bf h_{f_j(v)} } \ri)\;.
$$
We recall that in this formula the edges $e_j$ are the  edges incident to $v$,  oriented away from $v$ and counted starting from the stem of the cherry in counterclockwise order.  Similarly the faces are the incident faces (triangles) counted from the face containing the stem. 

Separating the  contributions to $\omega_v$ into the types $(z,z)$, $(z,x)$ and $(x,x)$ we come to  (\ref{omegav}).

\noindent {\bf Contributions of the ``face'' vertices $p_f\in V(\Sigma)$.}
For each $f\in F(T)$ we have a contribution $\omega_f$ as in \eqref{omegaf} in terms of the variables ${\bf x} = {\bf x}_f$, given by  Prop. \ref{propomegaf}. 

\noindent{\bf Contribution of the cherries.}
For each cherry attached to the vertex  $v\in V(T)$   the local monodromy $M_v^0$ as well as the  diagonalizing matrix $C_v$ are lower--triangular. 
The contribution  to the form $\Omega(\Sigma_{FG})$  of the point $q_v$ (where the stem meets the cherry attached to the vertex $v$, see Fig. \ref{figtrian}) can be computed to give (since $C_v$ is triangular) 
$$
 2 \tr  \left((C_v)^{-1} \d C_v \wedge \Lambda_v^{-1} \d \Lambda_v 
\right)
=
2\sum_{j,k=1}^{n-1}\tr(\alpha_j {\mathrm h}_k)\d  \rho_{v;k}  \wedge  \d \mu_{v;j} 
=2n \sum_{j=1}^{n-1} \d \rho_{v;j} \wedge \d \mu_{v;j}
$$
where we have used that $\tr(\a_j {\mathrm h}_k) = n \delta_{jk}$ \eqref{rootprods}.
\QED

\subsection{ The form $\Oh $ via Fock-Goncharov coordinates} 

Here we use the invariance of the form $\Omega(\Sigma)$ under the graph transformations to  transform the graphs
$\Sigma_{AM}$ and $\Sigma_{FG}$ to the same graph,  which we denote by $\wh\Sigma$ (Fig.  \ref{figAM}). This will lead
to an expression of the form $\Oh $ via Fock-Goncharov coordinates.

\begin{theorem}
\label{thm5}
Let the  matrices on the edges of  the graph $\wh \Sigma$ shown in Fig. \ref{figAM} be obtained by standard transformations from the edge matrices on the graph $\Sigma_{FG}$. Then the  form 
 ${2 \Oh}=   \Omega(\wh \Sigma)$ 
coincides with the form $\Omega(\Sigma_{FG})$ given by expression (\ref{ototal}).
\end{theorem}
{\bf Proof.} By an obvious sequence of standard transformations the graph $\Sigma_{AM}$ shown in Fig. \ref{figAMzip} can be transformed to the graph $\wh\Sigma$ shown in Fig. \ref{figAM}.  Namely, we first merge the $g$ pairs of vertices $s_\ell, \wt s_\ell$ into a vertex $\sigma_{\ell}$, then zip each pair 
of  corresponding closed edges together. On the loop edge obtained by zipping one gets the matrix $P_{2\ell} P_{2\ell-1}^{-1}  = M_{\beta_\ell}^{-1}$.
Then we merge all the vertices $\sigma_\ell$ with the basepoint $z_0$ and thus conclude that  $\Omega( \Sigma_{_{AM}})=\Omega(\wh \Sigma)$. On the other hand we have seen in Theorem \ref{thmWAM} that 
$2\mathcal W = \Omega(\Sigma_{_{AM}})$ and hence 
$2\mathcal W$ 
 also equals to $\Omega(\wh \Sigma)$.

The same form  $\Omega(\wh\Sigma)$ is also equal to the form $\Omega(\Sigma_{FG})$ since the graph $\wh\Sigma$, together with matrices on its edges, can be obtained  by a sequence of transformations from the graph $\Sigma_{FG}$. Namely, by a sequence of edge contractions, we glue all vertices $v\in \mathbf V(\Sigma_0)$ to a single vertex $z_0$. We can then move all the cherries  to  the same region bounded by two consecutive edges at $z_0$. Finally we zip the edges so that we end up with a minimal number ($2g$) as in Fig. \ref{figAM}. By the results of Section \ref{secmoves} the two-form $\Omega$ remains invariant. \QED
Theorem \ref{thm5} shows that the Fock--Goncharov coordinates provide log--canonical coordinates for the extended Goldman symplectic form.
\begin{figure}
\begin{center}
\includegraphics[width=0.7\textwidth]{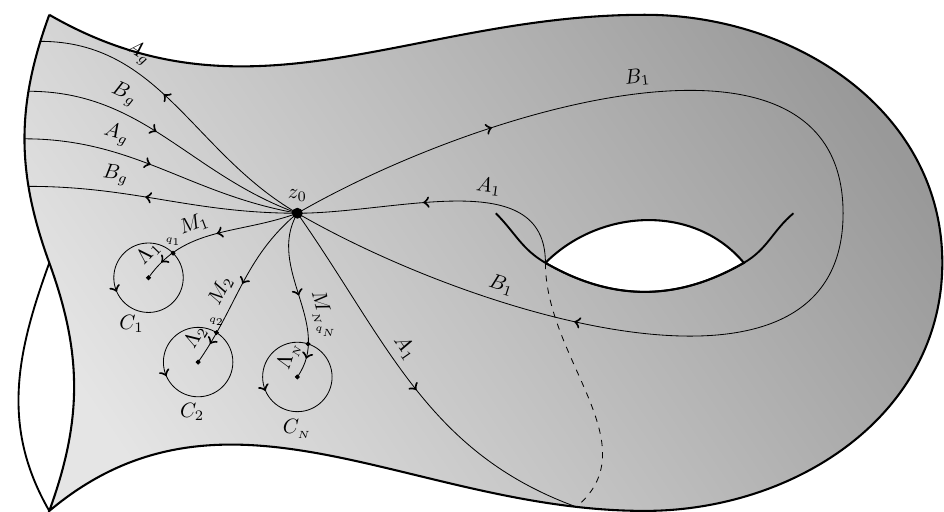}
\end{center}
\caption{The graph $\wh\Sigma$ and the jump matrices on its edges.}
\label{figAM}
\end{figure}

\section{The Poisson structure: extension of the Fock--Goncharov quiver} 

Here we discuss the Poisson bracket which inverts the nondegenerate symplectic form $\Oh$ (\ref{defCh}).
The actual proof will be given for the $SL(2)$  case, for $n>2$ the Poisson bracket $\{\cdot,\cdot\}$ described below was 
confirmed by extensive computer experiments.   This bracket is an extension of the Goldman bracket $\{\cdot,\cdot\}_{_G}$ which can be described in terms of Fock-Goncharov coordinates  by an appropriate quiver on the underlying Riemann surface \cite{FG}; the equivalence of the Goldman bracket and the Fock-Goncharov bracket was addressed in the recent paper \cite{ZheSun}. To describe the bracket  $\{\cdot,\cdot\}$  we introduce the 
quiver ${\mathscr Q}$ which is an extension of the Fock-Goncharov quiver  ${\mathscr Q}_{FG}$ obtained
by addition of  nodes  corresponding to the toric variables $\rho_{v;j}, \ j=1,\dots, n-1$.

Following \cite{FG} we consider  triangulation $\Sigma_0$ and perform a subtriangulation of each of the triangles into sub-triangles; the internal vertices are labelled by three indices $a,b,c\geq 1$ such that $a+b+c = n$. The corner attached to the edge $\mathcal E_f^{(1)}$ is the corner labelled 
$(n,0,0)$ , the corner attached to $\mathcal E_{f}^{(2)}$ is $(0,n,0)$ and the one attached to $\mathcal E_f^{(3)}$ is $(0,0,n)$. 
Then  black vertices and black arrows correspond to the (part of) the quiver ${\mathscr Q}_{FG}$, where
the nodes at the edge $e$  of the face $f$ correspond to variables with 2 indices. Namely, 
 the edge variables $\zeta_{e;1},\dots, \zeta_{e;n-1}$ between vertices $1$ and $2$ are labelled by triples of the form $(j,n-j,0)$; the variables  between vertices $3$ and $1$ carry the
  labels $(n-j,0,j)$;  finally, the variables  on the edge between $3$ and $2$ are labelled by triples of the type $(0,j,n-j)$.  The edge variables appear also in the neighbouring triangle. The internal nodes of the face $f$ are labelled by three indices $a,b,c\geq 1$ such that $a+b+c = n$ and carry the variables $\xi_{abc}$.

In the faces of $\Sigma_0$ which do not contain any cherry the extended quiver ${\mathscr Q}$ is defined to coincide with 
${\mathscr Q}_{FG}$.
The nodes which are added to ${\mathscr Q}_{FG}$ to get the full quiver ${\mathscr Q}$ carry the toric variables $\rho_{v;j},\ \ j=1,\dots, n-1$, $v\in {\bf V}(\Sigma_0)$; these {``toric''} nodes appear if there are cherries inside of a given face $f$ { and are shown in red color (in the online version)}.  These {toric} nodes are placed above the face $f$ and project normally to the sub-triangles on the closest edge to the corresponding stem of the cherry as shown  in Fig.  \ref{quiverFG}. Depending on the number of cherries within the same triangle $f\in \mathbf F (\Sigma_0)$, the quiver takes one of three forms shown in Fig.  \ref{quiverFG}.  The main example 
is when only one cherry is {situated } in the face $f$; this is possible unless the number of cherries exceeds the number of faces of $\Sigma_0$, which  is the case only for $g=0$, $n=3$.

\begin{figure}
\begin{center}
\includegraphics[width=0.98\textwidth]{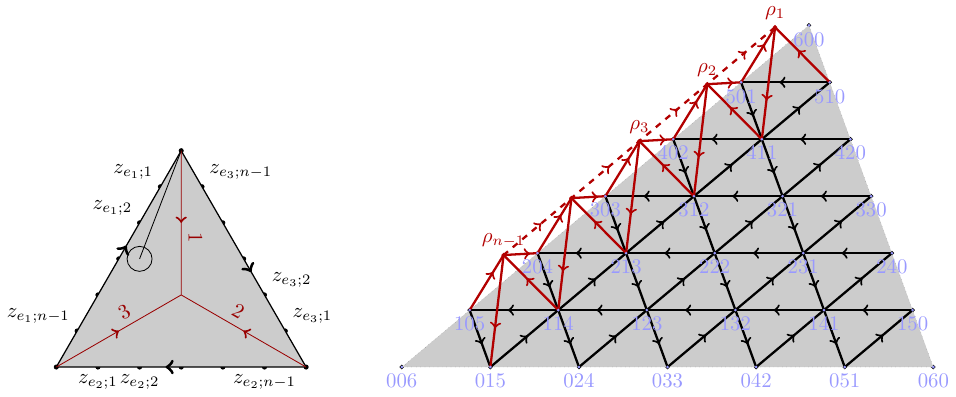}
\\
\includegraphics[width=0.98\textwidth]{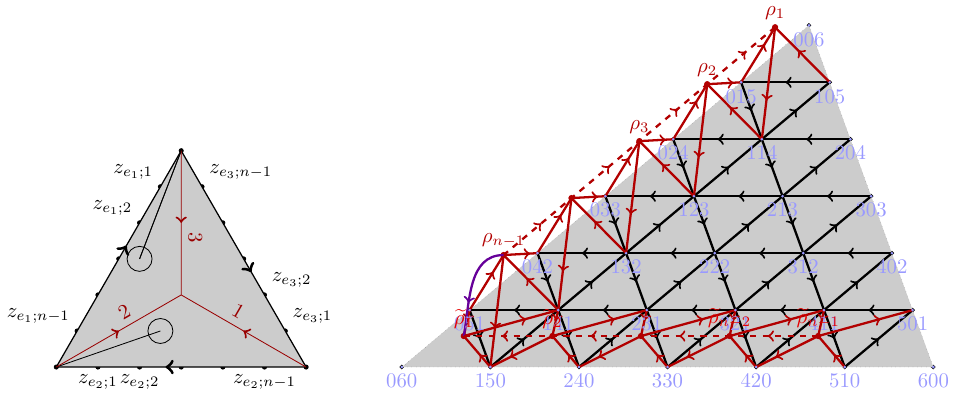}
\\
\includegraphics[width=0.98\textwidth]{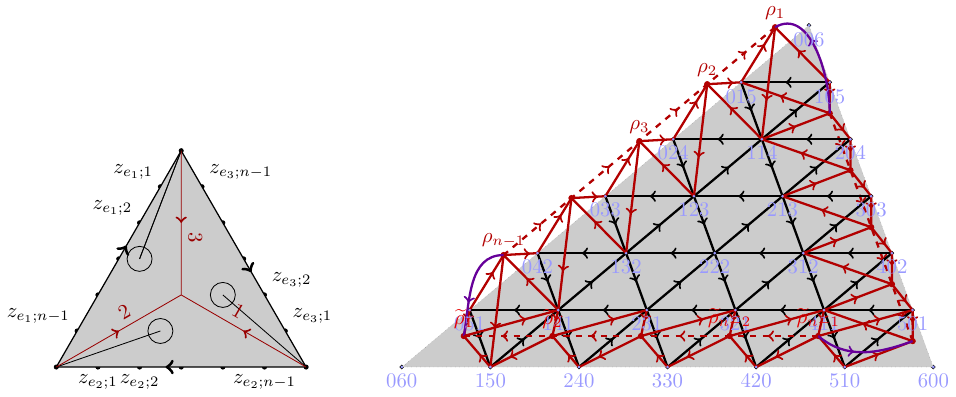}
\end{center}
\caption{The extended Fock-Goncharov quiver for a triangle with cherry  for the Poisson brackets multiplied with  $n^2$. On the left the positioning of the cherry relative to the numbering of the internal edges is shown. 
On the right we show the corresponding quiver (in pseudo--3d).  The dashed line means that the coefficients equals to  $\frac 1 2$, while all other coefficients are equal to  $1$ according to the indicated orientations. For a triangle without cherry, the picture is the same without the { toric} (raised) nodes and related arrows.} 
\label{quiverFG}
\end{figure}

Now we are in a position to  formulate the following 
\begin{conjecture}
Denote by $\sigma, \sigma'$ the logarithm of any two variables associated to two nodes $p,p'$ of the quiver indicated in Fig. \ref{quiverFG}. Then the Poisson bracket inverse to the extended Goldman symplectic form  $ \Oh $
is given by
\be
n^2 \{\sigma, \sigma'\} = \epsilon_{p,p'}
\la{PBnodes}
\ee
where  {$\epsilon_{p,p'} = 1$ if the arrow goes from $p$ to $p'$ and $-1$ viceversa}  or $\pm \frac 12$ if the arrow is a dashed one.
In particular, the brackets between Fock-Goncharov variables are given by the original quiver from \cite{FG}.
\end{conjecture}

Notice that part of this conjecture can be considered as a  theorem: the fact that the brackets between variables attached to two black nodes (i.e. the original Fock--Goncharov coordinates) are the same as the ones give by the quiver of \cite{FG}, follows 
from \cite{ZheSun,SZ}.
The conjecture is in fact a rigorous theorem for the $SL(2)$ case; we provide an essentially complete proof in the next section.  For the case of $SL(3)$ the direct proof is also possible: it consists of a lengthy verification along the lines of the  $SL(2)$ proof. Namely, one can  verify  that for each variable $\sigma_a$ we have $
\mathbb P \le(\Oh\le(\frac  \pa{\pa  \sigma_a}\ri) \ri)=$
$= - \frac {\pa}{\pa \sigma_a}$
 (or, equivalently, that $\Oh(\mathbb P ( \d \sigma_a)) =- \d\sigma_a$). Here  $\mathbb P$ denotes the Poisson tensor corresponding to the bracket \eqref{PBnodes}, and  $\Oh$ is viewed as a map from the tangent to the co-tangent space and viceversa for $\mathbb P$.  The proof for the $SL(n)$ case could clearly follow these ideas, but the combinatorics of the indices becomes quickly unwieldy. 

We notice the following structure of the bracket (\ref{PBnodes}). For each case except $g=0$, $n=3$ one can choose 
the positions of cherries such that in each face of $\Sigma_0$ there is no more than one cherry (Fig. \ref{quiverFG}, upper quiver). Then the toric variables $\rho_{v;j}$ commute with $\rho_{v';j}$ unless $v'=v$. At a given vertex $v$ we have 
$$
n^2\{\rho_{v;j},\rho_{v;j+1}\}=-\f{1}{2}\;,\hskip0.8cm j=1,\dots,n-2
$$
and all other brackets vanish.
The Poisson brackets between the toric variables and Fock-Goncharov variables can be seen from the  quiver shown
in Fig.\ref{quiverFG}, upper pane. Each $\rho_{v;j}$ has non-vanishing Poisson bracket with no more than four Fock-Goncharov variables, and these non-vanishing brackets are equal to $\pm 1$.

\section{Symplectic potential for $\Omega(\Sigma_{_{FG}}) $ and its independence from  the ciliation}
Since the  two-form    $\Omega(\Sigma_{_{FG}}) $   \eqref{ototal} has constant coefficients, one can write a corresponding primitive explicitly. Amongst the possible primitives, we select the following definition, which, as we show below, is invariant under the choice of ciliation: 
\begin{definition}
\label{deftheta}
The symplectic potential $\theta(\Sigma_{_{FG}})$ is defined by
\be
\label{thetaSLn}
\theta(\Sigma_{_{FG}}) = \sum_{v \in {\bf  V}(\Sigma_0)} \theta_v +  \sum_{f \in {\bf  F}(\Sigma_0)} \theta_f + 
n \sum_{v\in {\bf V}(\Sigma_0)} \sum_{j=1}^{n-1}\le(\rho_{v;j} \d \mu_{v;j} - \mu_{v;j} \d \rho_{v;j}\ri)
\ee
where, following \eqref{omegav}, we define
\begin{align}
\label{thetav}
 2\theta_v = &
 \sum_{e'\prec e \perp v}   \sum_{i,j=1}^{n-1}\CM_{ij} ( {\zeta}_{e'; i}\d  {\zeta}_{e;j}-   {\zeta}_{e;j}  \d {\zeta}_{e'; i})
 \nn\\
&+\sum_{f\prec e\perp v} \sum_{a+b+c =n} \sum_{\ell=1}^{n-1} \CM_{f(v),\ell}
 (\xi_{f;abc}   \d \zeta_{e;\ell}  - \zeta_{e;\ell}  \d \xi_{f;abc})
\nn \\
&+
\sum_{e\prec f\perp v} \sum_{a+b+c =n}\sum_{\ell=1}^{n-1}  \CM_{f(v),\ell} 
( \zeta_{e;\ell}  \d \xi_{f;abc} -    \xi_{f;abc}\d \zeta_{e;\ell})
\nn \\
&+\sum_{ f' \prec  f\perp v }  \sum_{a+b+c=n\atop a'\!\!+b'\!\!+c'\!=n}\CM_{f'(v),f(v)}  
({\xi}_{f'; a'b'c'}   \d  {\xi}_{f;abc} -   {\xi}_{f;abc} \d {\xi}_{f'; a'b'c'})
\end{align}
and, following \eqref{omegaf} we define
\begin{align}
\label{thetaf}
2\theta_f
=\sum_{i + j + k = n\atop 
i' + j' + k' = n}
F_{ijk; i'j'k'} (  \xi_{f; ijk}  \d \xi_{f; i'j'k'} - \xi_{f; i'j'k'}\d \xi_{f; ijk}  )
\end{align}
with the coefficients defined in \eqref{coefface}.
\end{definition}

Obviously $\d \theta(\Sigma_{_{FG}}) = \Omega( \Sigma_{_{FG}})$. 
The terms $\theta_f$ do not depend on the positions of cherries, but the terms $\theta_v$ do.
However, we have  the following theorem.

\begin{theorem}
\label{nohair}
The form $\theta (\Sigma_{_{FG}})$ given by \eqref{thetaSLn} is independent of the position of the cherries.
\end{theorem}
{\bf Proof.}
The proof is an elementary computation. We provide details in the simplest case where the vertex $v$ has only two edges in $\Sigma_{0}$ (and hence also two edges $\mathcal E$). 
The vertex $v$ is on the boundary of two triangles $f_1,f_2$ and we choose the marking of the edges $\mathcal E_{f}^{(j)}$ so that on both faces the vertex $v$ is attached to $\mathcal E_{f_1}^{(1)}$ and $\mathcal E_{f_2}^{(1)}$. In these notation we have  $f_1(v)=f_2(v)=1$. 
Let 
\be
&\eta_{1;j} = \sum_{b+c=n-j} \xi_{f_1;jbc}; \ \ \  \  \eta_{2;j} = \zeta_{e_1;j} ;\nn
\\
&\eta_{3;j} = \sum_{b+c=n-j} \xi_{f_2;jbc}; \ \ \ \   \eta _{4;j} = \zeta_{e_2;j},  \ \ \ \   j=1,\dots, n-1\; . 
\ee
The $(n-1)$--dimensional vectors ${\bs \eta }_\ell$, $\ell=1,2,3,4$ correspond to the four edges incident to $v$ (two ``red'' and two ``black'', see Fig. \ref{figtrian}), counted  in counterclockwise order starting from the initial position of the cherry. 

Let us denote 
\be
\langle j,k\rangle  := \frac{{\bs \eta }_j^\top  \CM \d {\bs \eta}_k - {\bs \eta}_k^\top  \CM \d {\bs \eta}_j}2, \hskip0.7cm  j,k\in \{1,2,3,4\}\;.
\ee 
Then the contribution of the vertex $v$ to $\theta(\Sigma_{_{FG}})$ in \eqref{thetaSLn} is 
\be
\label{before}
\theta_v = \langle 1,2\rangle  + \langle 1,3\rangle  + \langle 1,4\rangle  + \langle 2,3\rangle  + \langle 2,4\rangle   + \langle 3,4\rangle \; .
\ee
When moving the cherry in the other triangle, the order of the incident vertices is rotated cyclically twice and becomes $\{3,4,1,2\}$, so that 
\be
\label{after}
\widetilde \theta_v = \langle 3,4\rangle  + \langle 3,1\rangle  + \langle 3,2\rangle  + \langle 4,1\rangle  + \langle 4,2\rangle  + \langle 1,2\rangle\; .
\ee
Keeping in mind that $\langle j,k\rangle  = -\langle k,j\rangle $, the difference between the two expressions \eqref{before} and \eqref{after} is 
\be
\widetilde \theta_v - \theta_v  = 2\big(
\langle 2,4\rangle  + \langle 1,3\rangle   + \langle 2,3\rangle   + \langle 2,4\rangle 
\big)\;.
\label{deltathetav}
\ee

Now we inspect how the  toric variables $\bs\rho$ change. 
We have seen in Proposition \ref{movcher}  that when we  move an edge from the left to the right of the cherry, the matrix  $C_v$ is multiplied from the left by $J_1^{-1}$.  According to our convention that the cherry is always to the right of an edge $\mathcal E_f^{(j)}$  
(see Fig. \ref{figtrian}), we should check what happens to the symplectic potential $\theta(\Sigma_{_{FG}})$ \eqref{thetaSLn} when moving an ``$A$-edge'' and an ``$S$-edge'' from the left to the right of the cherry. 

  Under such a move the monodromy is conjugated by $J_1 = A_1 S_1$, which is lower triangular like $C_v$ and $M_v^{(0)}$. Thus the diagonal entries of $M_v^{(0)}$ are preserved and since $\wt C_v = S_1^{-1} A_1^{-1} C_v$,   the diagonal part of $C_v$ is multiplied by  the inverse of the diagonal part of $A_1 S_1$. Using \eqref{defA1}, a direct inspection reveals that 
\be
(A_1 S_1)^D = P  {\bf x}_{f_1}^{{\bf h_1}}   {\bf z}_{e_1}^{\bf h}  P\;.
\ee
Thus the  variables $\rho_{v;j}$ undergo a shift
\be
\widetilde \rho_{v;j} = \rho_{v;j }  - \sum_{ b,c\atop  b + c =j} \xi_{jbc} - \zeta_{e;n-j}, \ \ \ \ j=1,\dots, n-1\; ,
\ee
or, equivalently,
 \be
\label{tilderho}
\widetilde {\bs \rho} = P ({\bs \eta}_{1} + {\bs \eta}_2) P 
\ee
where $P $ here is the long permutation in $GL(n-1)$. 
The formula \eqref{muvj} can be written similarly as 
\be
{\bs \mu} =\frac 1 n P \CM \big( {\bs \eta}_1 + {\bs \eta}_2 + {\bs \eta}_3+ {\bs \eta}_4 \big) \;.
\ee
Thus contribution of the vertex $v$ in the last sum of formula \eqref{thetaSLn} can be written as 
$$
\d {\bs \rho}^\top \CM \big( {\bs \eta}_1 + {\bs \eta}_2 + {\bs \eta}_3+ {\bs \eta}_4 \big) -  {\bs \rho}^\top \CM \d \big( {\bs \eta}_1 + {\bs \eta}_2 + {\bs \eta}_3+ {\bs \eta}_4 \big)
$$
 (recalling that $P  \CM P  = \CM$). 
Now using \eqref{tilderho} we see that 
\be
n\le(\d ({\bs {\widetilde \rho}} - {\bs \rho}) ^\top {\bs \mu} -  ({\bs {\widetilde \rho}} - {\bs \rho}) ^\top \d{\bs \mu} \ri) = 
 - 2 ( \langle 1,3\rangle  + \langle 1,4\rangle  + \langle 2,3\rangle  + \langle 2,4\rangle ),
\ee
which cancels against \eqref{deltathetav}.\QED

\section{SL(2)}
\label{SL2mon}
 In the $SL(2)$ case the jump matrices on the oriented edges of $\Sigma$   (Fig. \ref{figtrian})  have the following expressions:
\begin{enumerate}
\item
 On each  edge $e$ which is inherited by $\Sigma$ from $\Sigma_0$  we define the jump matrix to be 
\be
S_{e}=\left(\ba{cc} 0 & 1/z_{e} \\ -z_{e} & 0 \ea\right) 
\la{Sijdef}
\ee
where $z_e\in \C$.
Note that $S_{e}^{-1} = -S_{e}$.
\item
The jump matrices on $\Ec_{k}^{(i)}$ do not contain any variables and are  given by 
$
A=\left(\ba{cc} 0 & -1 \\ 1 & -1 \ea\right)\;.
$

\item The jump matrix on the stem of the cherry attached to a vertex $v=v_j$ which has valence $q$ on $\Sigma_0$ (and valence $q+1$ on $\Sigma$) is chosen such that the total monodromy around $v$ is trivial due to \eqref{nomono}. Namely,

\be
M_v^0=(-1)^{\#_v}\left(\prod_{\ell =1}^q  A S (e_\ell)\right)^{-1}=\left(\ba{cc} m_v & 0 \\ \star & m_v^{-1} \ea\right)
\la{J0}
\ee
where $\#_v$ is the number of edges incoming to the vertex $v$. Therefore,
\be
m_v=(-1)^{\#_v}\prod_{e\perp v} z_e \;.
\label{m_j} 
\ee
The (local) connection matrix $C^0_{v}$ is lower--triangular and of the form:
\be 
\label{CR}
C^0_{v}=\left(\ba{cc} r_v & 0 \\ \star & r_v^{-1}  \ea\right).
\ee
\end{enumerate}

 In the  $SL(2)$ case  the face variables are absent and each edge carries a single variable, 
while the  eigenvalue  $m_v$ is  (up to a sign which is irrelevant in the expression of $\Omega(\Sigma_{_{FG}}) $) the product of the edge $z$--variables incident to $v$.

Then the general formula in Thm.  \ref{thmtoric} simplifies considerably to the following (we write the sum over vertices of the graph $\Sigma_0$, reminding the reader that the vertices of $\Sigma_0$ are in one-to-one correspondence with the $N$ punctures)
\be
\Omega(\Sigma_{_{FG}}) =2\sum_{v\in \mathbf V(\Sigma_0) } \le(\sum_{e,e'
\perp v\atop e'\prec e}  \d \zeta_{e'}\wedge \d \zeta_{e}
 +2 \sum_{e\perp v}  \d\rho_v \wedge \d \zeta_e \ri)\;.
 \la{omegaMlog}
\ee
The symplectic potential  \eqref{thetaSLn} takes the form: 
\be
 \theta(\Sigma_{_{FG}}) =\sum_{v\in \mathbf V(\Sigma_0) }\le(\sum_{e,e'\perp v\atop e' \prec e}(\zeta_{e'}   \, \d \zeta_{e}- \zeta_{e}  \d \zeta_{e'})
 + 2 \sum_{e\perp v} ( {\rho_v \d\zeta_e} - \zeta_e\, \d \rho_v ) \ri).
 \la{sympotM}
\ee
The choice of  $\theta_\Mcal$ depends on the choice  of  triangulation $\Sigma_0$.  As well as the general $SL(n)$ case, the $SL(2)$  potential $\theta_\Mcal$ transforms in a nontrivial way  under the change of triangulation; 
this transformation is discussed in the next section.

\subsection{Extended (nondegenerate) Poisson structure}
\label{secpoiss}
It is possible to write explicitly the Poisson structure, i.e., the inverse transpose  of the matrix of coefficients of $\Omega(\Sigma_{_{FG}})$. The idea is to observe the coincidence of the restriction of  $\Omega(\Sigma_{_{FG}})$ to the symplectic leaves with the Kontsevich symplectic form associated to the combinatorial model of $\mathcal M_{g,N}$ and use results of \cite {BK}. 

Recall that the vertex variables $\rho_v$ are associated to the stem of the cherry; this belongs to a particular triangle $f\in {\bf F}(\Sigma_0)$ of the triangulation $\Sigma_0$. In this way we can unambiguously declare that $v\in f$. This way every vertex ``belongs'' to a certain unique triangle $f$. Depending on how we have chosen the positions of the cherries, some faces may ``contain" zero, one, two or all three vertices.

Now we can state the theorem:
\begin{theorem}
The Poisson tensor induced by the symplectic structure 
 $ \Oh ={ \frac 1 2} \Omega(\Sigma_{_{FG}})$
  \eqref{omegaMlog}  is given by 
\be\label{Poiss}
\mathbb P   & =  \sum_{f\in {\bf F}(\Sigma_0) } \mathbb P_f
\ee
with
\be
\label{4Poiss}
4  \mathbb P_f= & \sum_{1\leq i < j \leq 3}\!\!\! (-1)^{i-j} \frac {\pa}{\pa \zeta_i} \wedge \frac {\pa}{\pa \zeta_{j}} 
\ee
$$
+
 \sum_{j=1}^3 \sum_{{\ds v: \atop  {v}_{|} \in f}}  (-1)^ {(v,e_j)}  \frac {\pa}{\pa \zeta_{j}} \wedge \frac {\pa}{\pa \rho_v } 
+ \!\!\sum_{v\prec v' \in f:\atop v_|, v'_|\in f}\!\!\! (-1)^{\sharp_f (v,v')} \frac {\pa}{\pa \rho_{v} } \wedge \frac {\pa}{\pa \rho_{v'}}
$$
where  $\zeta_{1,2,3}$ are the three edge variables of the triangle $f$ enumerated counterclockwise (starting from an arbitrarily chosen one)  and  $(v,e_j) = 1$ if $e_j$ is incident to $v$ and zero otherwise, and $\sharp_f(v,v') = 1$ if $v$ is the immediate predecessor of $v'$ along the boundary of the triangle $f$ (in positive direction), and zero otherwise. 
\end{theorem}   
\br\rm
We emphasize that here the symbol $ {v}_{|} \in f $ means that the corresponding cherry at $v$ belongs to the triangle $f$ and not only that $v$ lies on the boundary of $f$. 
\er
{\bf Proof.} The proof is rather direct and here we only provide a sketch.  For brevity of notation we set $\wh \Omega:= \Omega(\Sigma_{_{FG}} ) = -2\Oh$, with $\Omega(\Sigma_{_{FG}})$ defined in \eqref{omegaMlog}. 
Consider first a vertex and its associated variable $\rho_v$; we will interpret $\wh\Omega$ and $\mathbb P$ as maps from the tangent to the co-tangent spaces  and viceversa. 
For brevity we will write $\d v$ for $\d \rho_v$, $\partial_v$ for $\partial_{\rho_v}$,  $\d e$ for $\d \zeta_e$ and so on. 
Enumerate the edges incident at $v$ by $e_1,\dots, e_{n_v}$ (counterclockwise), starting from the first to the left of the stem (see Fig. \ref{figvert}). Denote by $\wt e_j$ the third edge in the face bounded by $e_j$ and $e_{j+1}$. 
\begin{figure}
\begin{center}
\begin{tikzpicture}[scale=2.2]
\draw [fill](0,0) circle [radius =0.03];
\coordinate (n) at (1,0);
\draw(0,0) to node[pos=0.7,above ]{$e_1$}(1,0);
\foreach \i/\L/\E in { 20/$\wt e_1$/$e_2$,  50/$\wt e_2$/$e_3$, 80/$\wt e_3 $/$ $, 120/$ $/$ $, 180/$ $/$ $, 230/$ $/$ $, 260/$ $/$ $, 300/$ $/$ $, 320/$ $/$e_{n_v}$ }{
\draw (0,0) to node[pos=0.7, above] {\E} (\i:1) coordinate(c) ;
 \draw (c) to node[pos=0.5,right] {\L}(n); 
\coordinate(n) at (c);
};
\draw(c) to  node[pos=0.5,right] {$\wt e_{n_v}$} (1,0);
\draw[ |-> , dashed] (0.3,0) arc(0:280:0.3);
\end{tikzpicture}
\hspace{1cm}
\begin{tikzpicture}[scale=2.2]
\coordinate(v1) at (-0.5,0);
\coordinate(v2) at (0.5,0);
\coordinate(w1) at ($(0.5,0) + (120:1)$);
\coordinate(w2) at ($(-0.5,0) + (-60:1)$);
\draw(v1)to (w1) to (v2);
\draw(v1)to (w2) to (v2);
\coordinate(t1) at ( $(v1) + (150:0.4)$);
\draw (t1) circle[radius=0.1];
\draw (v1) to (t1);
\node [above]  at ($ (v1) + (0,0.16)$) {$v$}; 
\node [below]  at ($ (v2) + (0,-0.15)$) {$v'$} ;  
\coordinate(t2) at ( $(v2) + (-10:0.4)$);
\draw (t2) circle[radius=0.1];
\draw (v2) to (t2);

\draw [fill](v1) circle [radius =0.03];
\draw [fill](v2) circle [radius =0.03];
\draw (v1) to node [pos=0.4,above] {$e_{\ell} = e$} node [pos=0.6, below] {$e= e_{\ell'}$} (v2);
\coordinate (n) at (v2);
\foreach \i/\L/\E in {60/$ $/$ $, 120/$ $/$ $, 180/$ $/$ $, 230/$ $/$ $, 260/$ $/$ $, 300/$ $/$ $ }{
\draw (v1) to node[pos=0.7, above] {\E} ($(v1)+ (\i:1)$)  coordinate(c) ;
 \draw (c) to node[pos=0.5,right] {\L}(n); 
\coordinate(n) at (c);
};
\coordinate(n) at (v1);
\foreach \i/\L/\E in {120/$ $/$ $, 80/$ $/$ $, 20/$ $/$ $, -50/$ $/$ $, -120/$ $/$ $ }{
\draw (v2) to node[pos=0.7, above] {\E} ($(v2)+ (\i:1)$)  coordinate(c) ;
 \draw (c) to node[pos=0.5,right] {\L}(n); 
\coordinate(n) at (c);
};

\draw[ |-> , dashed] ($(v1) + (0.2,0)$)  arc(0:280:0.2);
\draw[ |-> , dashed] ($(v2) + (0.2,0)$)  arc(0:280:0.2);
\end{tikzpicture}

\end{center}
\caption{}
\label{figvert}
\end{figure}
Then 
\be 
\wh\Omega (\pa_{_v})=- 4\sum_{j=1}^{n_v} \d {e_j}  \;.
\ee
Now {let us check that } $4\mathbb P(\wh\Omega(\pa_{_v}))=-8\pa_v$ {(or, equivalently, that $\mathbb P(\Oh(\pa_v))=-\pa_v$)}. 
The action of the last sum in \eqref{4Poiss} on   $\wh\Omega(\pa_{_v})$   vanishes; the second term gives 
\be
\le(-  \pa_{e_1} \wedge  \pa_{_v}  -  \pa_{e_{n_v}} \wedge \pa_{v} \ri) {\big \lrcorner} \le(-4\sum_{j=1}^{n_v} \d {e_j}\ri)  =-8 \pa_{_v}\;.
\ee
{The total contribution (summing over all triangles $f$) of } the first sum  in \eqref{4Poiss} on  $\wh\Omega(\pa_{_v})$  also equals  zero. Indeed only the faces incident to $v$ are involved and  the result is a telescopic sum 
$$
\sum_{j=1}^{{n_v}} \le( \pa _{\wt e_j}  \wedge \pa_{e_j}- 
\pa_{e_{j+1}} \wedge \pa_{e_j}  + 
 \pa_{e_{j+1}} \wedge \pa_{\wt e_j} 
\ri) {\big \lrcorner} \le(-4\sum_{\ell=1}^{n_v} \d e_\ell\ri) 
$$
$$
=
-4\sum_{\ell=1}^{n_v} \le(
\pa_{\wt e_\ell} - \pa_{e_{\ell+1}} + \pa_{e_{\ell-1}} - \pa_{\wt e_{\ell-1}}\ri) =0 \;;
$$
$$
e_{n_v+1}\equiv e_1\,,  \ \ \ e_{0} \equiv e_{n_v}\,,  \ \ 
\wt e_{n_v+1}\equiv \wt e_1\,,  \ \ \ \wt e_{0} \equiv \wt e_{n_v}\;.
$$
Consider now an edge $e$ joining $v, v'$; let $e_1,\dots, e_{n_v}$ be the enumeration of incident edges at $v$ and similarly $e_1', \dots, e'_{n_{v'}}$ be the enumeration of edges at $v'$. The edge $e$ is the edge number $\ell$ at $v$ and the edge number $\ell'$ at $v'$. Then 
\be
\wh\Omega(\pa_{_e}) = 2 \d  v + 2 \d {v'} -  \sum_{j=\ell+1}^{n_v} \d e_j   + \sum_{j=1}^{\ell-1} \d e_j  -
 \sum_{k=\ell'+1}^{n_{v'}} \d e'_{k}  + \sum_{k=1}^{\ell'-1} \d e'_{k}\;  .
\ee
Now we  contract the above with $\mathbb P$: after a somewhat lengthy computation one finds $\mathbb P(\Oh(\pa_e)) ={- \pa_e}$. 
\QED

\begin{theorem}
The Goldman Poisson tensor on the subspace of the $SL(2)$ character of $n$-punctured Riemann surfaces of genus $g$  variety covered by 
shear coordinates associated to triangulation $\Sigma_0$ is given by 
\be
\label{PBDirac}
{\mathbb P}_G =  -\f{1}{4}
 \sum_{f\in {\bf F}(\Sigma_0) } \left(\f{\p}{\p \zeta_{e_1}}\wedge  \f{\p}{\p \zeta_{e_2}}+\f{\p}{\p \zeta_{e_2}}\wedge  \f{\p}{\p \zeta_{e_3}}
+\f{\p}{\p \zeta_{e_3}}\wedge  \f{\p}{\p \zeta_{e_1}}\right)
\ee
where $e_1$ $e_2$ and $e_3$ are the edges (counted counter-clockwise) which form the boundary of the face $f$ of $\Sigma_0$.

The inverse of the Poisson tensor (\ref{PBDirac}) on the symplectic leaf $\mu_j=const$ is given by the Goldman symplectic form
\be
\la{omegamu}
\Oh_{G}=\sum_{v\in \mathbf V(\Sigma_0) } \sum_{e,e'
\perp v\atop e'\prec e}  \d \zeta_{e'}\wedge \d \zeta_{e}\;.
\ee
\end{theorem}

{\bf Proof}. 
First, one verifies directly that for the bracket \eqref{Poiss} the expressions $\mu_v= \sum_{e\perp v} \zeta_e$ Poisson-commute with all the variables $\zeta_e$ for all $v\in {\bf V}(\Sigma_0), \ e\in {\bf E}(\Sigma_0)$, and $\{  \rho_{v'},\mu_v\}  = \frac 12\delta_{vv'}$. This is a simple exercise which we do not report in detail.

The quotient of $\CVh$ by the toric action of multiplication of each $C_j$ on the right by a diagonal matrix 
(this quotient amounts to arbitrary translations in the variables $\rho_v$)
can be naturally mapped   to the open-dense set $\mathcal M$ of the character variety. Thus the push--forward of the Poisson tensor \eqref{Poiss} under this  quotient  map is obtained simply by removing the terms containing the derivatives with respect to $\rho_v$'s. This produces  the expression \eqref{PBDirac}. The Casimirs are thus the $\mu_v$'s written above.

On the other hand, the restriction of the Poisson tensor  ${\mathbb P}_G$ \eqref{PBDirac} to the symplectic leaves $\{\mu_v=$constant$\}$ coincides with  the Dirac reduction of the Poisson tensor \eqref{Poiss} to the level sets $\{\mu_v = $const, $\rho_v=$const$\}$. The Poisson tensor obtained via this Dirac reduction 
 is the same as the inverse of the restriction of the two--form $\Omega(\Sigma_{_{FG}})$ in \eqref{omegaMlog} to $\{\mu_v = $const, $\rho_v=$const$\}$, which yields  \eqref{omegamu}. This proves the second statement of the theorem. 

To prove the first statement, namely, that \eqref{PBDirac} coincides with the Goldman bracket, we recall that the result of  \cite{AM} says that the form \eqref{omegamu}, being restricted to the leaves $\mu_v=$const, gives the symplectic structure on the symplectic leaf of the Goldman bracket. 
With this we have shown that: 
\begin{enumerate}
\item The bracket \eqref{PBDirac} has the same Casimirs as the Goldman bracket and hence the same symplectic leaves; 
\item It coincides with the Goldman bracket on each symplectic leaf. 
\end{enumerate}
These two facts imply that \eqref{PBDirac} coincides with the Goldman bracket.
\QED 

\begin{remark}\rm
Algebraically, the  Poisson tensor ${\mathbb P}_G$ (\ref{PBDirac}) 
coincides (up to a multiplicative constant) with  the Weil-Petersson Poisson tensor $P_{WP}$ given by the formula  (73) of the paper by V.Fock \cite{Fock1} {(given there without detailed proof; for the published proof we refer 
to App.B of the paper by L.Chekhov \cite{Chekhov07}). 
The expression for the Weil-Petersson symplectic form in terms of $\lambda$-length coordinates was given by R.Penner in Th.A.2 of \cite{Penner} (see also \cite{Penner1}); his expression is different from our formula (\ref{omegamu}) for
$\Omega_{\boldsymbol \mu}$  since it's written in terms of different coordinates (the coordinates $\zeta_e$ 
in the $SL(2,\R)$ case are logarithms of shear coordinates while Penner's formula is written in terms of $\lambda$-lengths). In case of Riemann surfaces with holes and bordered cusps the  formula 
analogous to (\ref{omegamu}) was  found recently and independently of this paper in  \cite{Chekhov20}.}
\end{remark}

\begin{figure}
\begin{center}

\includegraphics[width=0.5\textwidth]{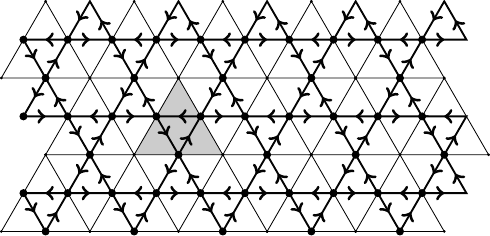}

\end{center}
\caption{The quiver for $SL(2)$ (thick), where we omit the nodes associated to the toric variables. The triangulation $\Sigma_0$ is shown in thin lines.
One face of $\Sigma_0$  is shaded for reference. }
\label{quiver}
\end{figure}

\begin{figure}
\begin{center}

\includegraphics[width=0.23\textwidth]{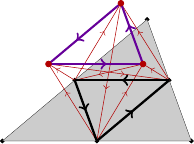} \ \ \ \ \ \ 
\includegraphics[width=0.23\textwidth]{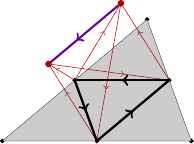} \ \ \ \ \ \ 
\includegraphics[width=0.23\textwidth]{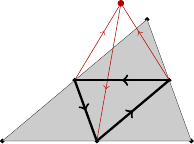}

\end{center}
\caption{The $SL(2)$ case of Figure \ref{quiverFG}. In the left pane we show   triangle containing three cherries and the corresponding augmented quiver. The  triangle shown in the center  contains two cherries, and the triangle shown in the right contains only  one cherry inside.  The quivers shown in the central and in the right panes  are obtained by deleting the corresponding toric nodes and all the incident edges.}
\label{SL2cherry}
\end{figure}

\br\rm
\label{FGrem}
Up to the overall factor of $4$, the Dirac Poisson bracket \eqref{PBDirac} can be expressed as the canonical Poisson bracket \cite{GekhtmanShapiro} associated to the following quiver: place a node on each edge $e\in {\bf E}(\Sigma_0)$ and triangulate the surface as shown in Fig. \ref{quiver} and Fig. \ref{SL2cherry}.  This appears to coincide with the Poisson structure introduced in \cite{FG}; in fact this coincidence is mentioned ibidem and on p.670 of \cite{FGbook}. We also point to \cite{ZheSun}, where the equivalence of the Goldman (degenerate, since it possesses Casimir functions) {\it Poisson structure} and the Poisson structure of \cite{FG} is shown for any $SL(n)$. 
\er
\subsection{Flip of an edge:  Rogers' dilogarithm as a generating function}
A triangulation can be transformed into any other by a sequence of "flips" of diagonal in
the quadrilateral formed by two triangles with a common edge, see Fig. \ref{flip}.
We are going to describe such a flip by assuming that the four  cherries attached to the vertices 
are placed as shown in Fig. \ref{flip}.
\begin{figure}
\begin{center}
\includegraphics[width=0.8\textwidth]{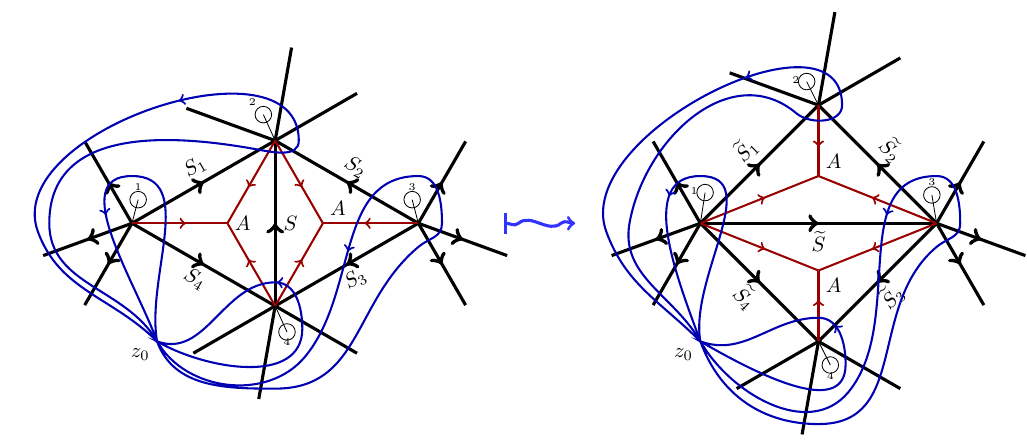}

\end{center}
\caption{
Transformation of edges and jump matrices under an elementary flip.}
\la{flip}
\end{figure}

{Assume that the base point is chosen on the external side of the edge 4 as shown in Fig. \ref{flip}.}
Then the assumption that all  the monodromies around the four
vertices of these triangles are preserved implies the following four equations:
\be
S_4 A S_1=\tilde{S}_4 A \tilde{S}A \tilde{S}_1\;, \hskip0.7cm
S_1^{-1} A S^{-1} A S_2^{-1}=\tilde{S}_1^{-1} A \tilde{S}_2^{-1}\;,
\la{SS1}
\ee
\be
S_2 A S_3=\tilde{S}_2 A \tilde{S}^{-1} A \tilde{S}_3\;,\hskip0.7cm
S_3^{-1} A S A S_4^{-1}=\tilde{S}_3^{-1} A \tilde{S}_4^{-1}\;.
\la{SS2}
\ee

In terms of variables $z$ and $z_j$ which parametrize the matrices $S$ and $S_j$, respectively, the equations (\ref{SS1}) and  (\ref{SS2}) have four different solutions. Two of these solutions take the form:
$$
\zt_1=\f{z}{(z^2+1)^{1/2}} z_1\;,\hskip0.5cm
\zt_2=-(z^2+1)^{1/2} z_2\;,
$$
\be
\zt_3=\f{z}{(z^2+1)^{1/2}} z_3\;,\hskip0.5cm
\zt_4= (z^2+1)^{1/2} z_4\;,\hskip0.5cm
\zt=\frac{1}{z}
\la{sol1}
\ee
and 
$$
\zt_1=-\f{z}{(z^2+1)^{1/2}} z_1\;,\hskip0.5cm
\zt_2=(z^2+1)^{1/2} z_2\;,
$$
\be
\zt_3=\f{z}{(z^2+1)^{1/2}} z_3\;,\hskip0.5cm
\zt_4= (z^2+1)^{1/2} z_4\;,\hskip0.5cm
\zt=-\frac{1}{z}\;.
\la{sol2}
\ee
Two other solutions are obtained by changing the determination of the square roots in the right-hand side of (\ref{sol1}) and (\ref{sol2}).

In all four cases the variables $\kappa_j=z_j^2$ transform as follows:
$$
 \kat_1=\f{\kappa}{\kappa+1} \kappa_1\;,\hskip0.5cm
\kat_2=(\kappa+1) \kappa_2\;,\hskip0.5cm
\kat_3=\f{\kappa}{\kappa+1} \kappa_3 \;,
$$
\be
\kat_4=(\kappa+1) \kappa_4\;,\hskip0.5cm
\kat=\frac{1}{\kappa}\;.
\la{kapkap}
\ee
The variables $\zeta_j=\f{1}{2}\log \ka_j$ are insensitive to the change of sign of $z_j$, and they transform as follows:
\bea
\zetat_1&=-\f{1}{2}\log(e^{2\zeta}+1) +\zeta+\zeta_1\;,\hskip0.5cm
\zetat_2= \f{1}{2}\log(e^{2\zeta}+1)+\zeta_2\;,
\nn\\
\zetat_3&=-\f{1}{2}\log(e^{2\zeta}+1) +\zeta+\zeta_3\;,\hskip0.5cm
\zetat_4=\f{1}{2}\log(e^{2\zeta}+1) +\zeta_4\;,
\nn\\
\zetat&=-\zeta\;.
\la{zeze}
\eea

Equations (\ref{zeze}) imply that all variables $\mu_v$ are preserved under the flip of an edge. Moreover, since not only the monodromy matrices themselves, but also the matrices $C_j$ 
are assumed  to be preserved under the move (since the cherries remain outside of the rectangle containing the flipping edge), the variables $r_j$ and $\rho_j=\log r_j$ are also preserved.

 Now we are going to compute the generating function of the edge flip.
 Introduce the   {\it Rogers dilogarithm}  $L$  which for  $x\geq 0$ is defined by the equality (we borrow this representation  from 
(1.9) of \cite{Nakanishi} and  refer also to \cite{Zagier} for more details):
\be
L\left(\f{x}{x+1}\right):=  \f{1}{2}\int_0^x \left\{ \f{\log(1+y)}{y}-\f{\log y}{1+y}\right\} {\rm d}y\;.
\la{dildef}\ee

Denote the new triangulation by $\wt{\Sigma}_{_{FG}}$ and  the corresponding symplectic potential (\ref{sympotM}) corresponding to the new triangulation 
 by  $\theta(\wt{\Sigma}_{_{FG}})$.
 
\begin{proposition}\la{potchange}
The symplectic potentials  ${\theta}(\wt \Sigma_{_{FG}})$ and $\theta(\Sigma_{_{FG}})$ are related as follows:
\be
{\theta}( \Sigma_{_{FG}})-{  \theta}( \wt \Sigma_{_{FG}}) = 2 \d \left[L\le(\frac {\ka}{1+\ka} \ri)\right]\;.
\la{ththt}
\ee
\end{proposition}
{\bf Proof.}
The  proposition can be verified by direct calculation using the definition (\ref{sympotM}) of the 
potential. The difference of contributions of the vertices $v_1,\dots,v_4$ to potentials 
${\theta}(\Sigma_{_{FG}})$ and ${\theta}(  \wt \Sigma_{_{FG}})$ equals
\beas
v_1:\hskip0.5cm  &  \frac 1 2\bigg(
\log \le(\frac {{\rm e}^{2 \zeta} }{{\rm e}^{2 \zeta}  + 1}\ri) \d\zeta_1
- \log({\rm e}^{2 \zeta}  + 1) \d \zeta_4
 + \zeta_4\d \log ({\rm e}^{2 \zeta}  +1 ) 
 \\
 &- \zeta_1 \d \log \le(\frac {{\rm e}^{2 \zeta} }{{\rm e}^{2 \zeta} +1} \ri)  +  \log ({\rm e}^{2 \zeta}  + 1) \d\zeta
 - \zeta \d \log({\rm e}^{2 \zeta}  + 1)   \bigg)
\\
v_2:\hskip0.5cm & 
\frac 1 2\bigg( -\log \le(\frac {{\rm e}^{2 \zeta} }{{\rm e}^{2 \zeta}  + 1}\ri) \d\zeta_1
+ \log({\rm e}^{2 \zeta}  + 1) \d\zeta_2
 -\zeta_2\d \log ({\rm e}^{2 \zeta}  +1 ) 
\\
 &  +\zeta_1 \d \log \le(\frac {{\rm e}^{2 \zeta} }{{\rm e}^{2 \zeta} +1} \ri)
 +  \log ({\rm e}^{2 \zeta}  + 1) \d\zeta - \zeta \d \log({\rm e}^{2 \zeta}  + 1)  \bigg)
\\
v_3:\hskip0.5cm  & 
\frac 1 2\bigg(\log \le(\frac {{\rm e}^{2 \zeta} }{{\rm e}^{2 \zeta}  + 1}\ri) \d \zeta_3
- \log({\rm e}^{2 \zeta}  + 1) \d\zeta_2
 +\zeta_2\d \log ({\rm e}^{2 \zeta}  +1 ) 
 \\
 & -\zeta_3 \d \log \le(\frac {{\rm e}^{2 \zeta} }{{\rm e}^{2 \zeta} +1} \ri)
  +  \log ({\rm e}^{2 \zeta}  + 1) \d\zeta -\zeta \d \log({\rm e}^{2 \zeta}  + 1)  \bigg)
\\
v_4:\hskip0.5cm  & 
\frac 1 2\bigg(-\log \le(\frac {{\rm e}^{2 \zeta} }{{\rm e}^{2 \zeta}  + 1}\ri) \d\zeta_3
+ \log({\rm e}^{2 \zeta}  + 1) \d\zeta_4
 - \zeta_4\d \log ({\rm e}^{2 \zeta}  +1 ) 
 \\
 & +\zeta_3 \d \log \le(\frac {{\rm e}^{2 \zeta} }{{\rm e}^{2 \zeta} +1} \ri)
   +  \log ({\rm e}^{2 \zeta}  + 1)\; \d\zeta -\zeta\, \d \log({\rm e}^{2 \zeta}  + 1)  \bigg)\;.
 \eeas

Summing up the above four contributions and taking into account the equation for the dilogarithm
we come to (\ref{ththt}). 
\QED

The Proposition (\ref{potchange}) means that the generating function of the symplectomorphism defined by the flip of an edge is given by
\be
G_{flip}=-2L\le(\frac {\ka}{1+\ka} \ri)\;.
\la{Gflip}
\ee
We notice that in the abstract setting of cluster algebras the Rogers' dilogarithm can be interpreted as an action of the  mutation \cite{Gekhtman}.

\subsection{$SL(2,\R)$: the dilogarithm circle bundle}

Let us  denote by $\CVh^{^{FG}}_{SL(2,\R)}$ the union of all open subsets of the real slice of $\CVh$ covered by the real coordinates $z_e, r_v$ for all possible triangulations:
the closure of $\CVh^{^{FG}}_{SL(2,\R)}$ is a  connected component of  $\CVh$ \cite{Kashaev}.
On this subset all $\kappa_e=z_e^2$ are real and positive.  The generating function (\ref{Gflip}) can then be used to define a circle bundle over the space $\CVh^{^{FG}}_{SL(2,\R)}$ canonically associated to the symplectic form (\ref{omegaMlog}) as we explain in this section. 
 
 Any two triangulations $T,T'$ can be connected  by a  finite sequence of elementary flips. 
 The  {\it exchange graph}  is  the abstract graph whose vertices are the triangulations and two triangulations are connected by an edge if they are related by an elementary flip. The exchange graph is known to be connected. 
 \bd
 The dilogarithm circle bundle $\mathscr D$ over $\CVh^{^{FG}}_{SL(2,\R)}$ is the circle bundle whose transition functions in the overlap of two triangulations $T,T'$,  differing by a flip of the edge $e$, are given by
 \be
 \label{transdilog}
 G_{T,T'} := \exp \le[-\frac {12}{i\pi }L\le(\frac{\kappa_e}{1+\kappa_e}\ri)\ri].
 \ee
 If two triangulations $T_1,T_2$ are connected by a sequence of elementary flips, the  transition function is the product of the elementary transition functions of the form (\ref{transdilog}).
 \ed
To verify the consistency of this definition one has to verify the cocycle conditions for the transition functions \eqref{transdilog}.
 
 We need two facts; the first is that  the Rogers' dilogarithm satisfies the following identities valid for  $x, y\in (0,1) \subset \R$:
\be
L(x)+ L(1-x)=\f{\pi^2}{6}\;,
\la{reflec}
\ee
\be
L(x)+L(y)+L(1-xy)+L\left(\f{1-x}{1-xy}\right)+L\left(\f{1-y}{1-xy}\right)= \f{\pi^2}{2} \;.
\la{fiveterm}
\ee
The second fact is that homotopy in the exchange graph is generated by square of flips and pentagon relations, see for example \cite{ChekhovFock}.

The relation \eqref{reflec} with $x=\frac {\kappa_e}{1+\kappa_e}$ guarantees that $G_{T,T'}G_{T'T}=1$ for any two neighbours $T,T'$ in the exchange graph.

The relation \eqref{fiveterm} implies that for any triangulations $T_1,\dots, T_5$ forming a pentagon in the exchange graph the transition functions satisfy the ``long'' cocycle condition $G_{T_1 T_2}\dots G_{T_5 T_1}=1$. Denoting the variables on the two internal edges flipped under the pentagon move by $\kappa_e$ and $\kappa_{e'}$ this cocycle condition gives rise to (\ref{fiveterm}) with 
$$
x=\f{\kappa_e}{\kappa_e+1}\;,\hskip0.7cm y= \f{\kappa_{e'}(\kappa_e+1)}{\kappa_{e'}(\kappa_e+1)+1}\;.
$$

\begin{remark}\rm 
We conclude by pointing out the difference between the real slice of the space $\CVh$ (\ref{defCVh}) and the symplectic leaf $\Mcal_{\Lambda}$ of the $SL(2,\R)$ character variety.
On the subspace $\Mcal_{\Lambda}^{FG}$ of $\Mcal_{\Lambda}$ the Goldman symplectic form $\Oh_G$ is given by
(\ref{omegamu}); thus we can choose 
 the symplectic potential as
$$
\theta_G =\f{1}{2}\sum_{v\in \mathbf V(\Sigma_0) }\sum_{e,e'\perp v\atop e' \prec e}(\zeta_{e'}   \, \d \zeta_{e}- \zeta_{e}  \d \zeta_{e'})\;.
$$
Assuming the positions of cherries (which define the ciliation of the graph) are chosen as in Fig. \ref{flip} the potential $\theta_G$ transforms in the same way as 
(\ref{ththt}): ${\theta}_G-\tilde{  \theta}_G =  \d \left[L\le(\frac {\ka}{1+\ka} \ri)\right]$ i.e. the Rogers' dilogarithm again gives the generating function of the flip.
However, while the potential $\theta(\Sigma_{FG})$ is invariant under the change of ciliation, this is not the case for the potential $\theta_G$: if  at the vertex $v$ the  cilium crosses the 
edge $e$ in the positive direction the potential ${\theta}_G$ transforms as
$$
{\theta}_G\to \theta_G - \d(\zeta_e \mu_v) \;.
$$
The dilogarithm circle bundle ${\mathcal D}$  is well-defined over the space  $\Mcal_{\Lambda}^{FG}$; presumably  its  first Chern class  is given by    $\Oh_G$ up to a linear combination of the forms representing the 
$\psi$-classes at the punctures, in analogy to  \cite{Mirz} (see also \cite{Wolpert}).
\end{remark}

\vskip 10pt
\noindent {\bf Acknowledgements.}  We thank L. Chekhov, S. Fomin, V. Fock, A. Goncharov, W. Goldman, R.  Kashaev and  M. Shapiro for  illuminating discussions and comments.  We thank the referee for useful comments.
The work of M. B. was supported in part by the Natural Sciences and Engineering Research Council of Canada (NSERC) grant RGPIN-2016-06660.
The work of D. K. was supported in part by the NSERC grant
RGPIN/3827-2015.
The completion of this  work was supported by the National Science Foundation under Grant No. DMS-1440140 while the authors were in residence at the Mathematical Sciences Research Institute in Berkeley, California, during the Fall 2019 semester
{\it Holomorphic Differentials in Mathematics and Physics}.

\appendix
\section{The form $\Oh $ in the $SL(3)$ case}

In the $SL(3)$ case the jump matrices on the oriented edges of the graph $\Sigma$
are chosen as follows.

\begin{enumerate}
\item
 On each  edge $e$ of $\Sigma_0$  the jump matrix is
\be
S_{e}= \le[
\begin{array}{ccc}
0 & 0 & \frac1{z_{e;1}^2 z_{e;2}}\\
0 & - \frac { z_{e;1}}{z_{e;2}} & 0 \\
 {z_{e;1}z_{e;2}^2} & 0 & 0 
\end{array}
\ri]
\ee
where $z_{e;i}  \in \C^*$, $i=1,2$.
Note that the transformation $S_{e}\to S_{e}^{-1}$ is equivalent to the interchange $z_{e,1}\leftrightarrow z_{e,2}$.

\item
The jump matrices on the edges $\Ec_{f}^{(1,2,3)}$ are  given by 
\be
A_f={x_f}\left(\ba{ccc} 0 & 0 & 1\\
0 & -1 & -1\\
x_f^{-3} & x_f^{-3}+1 & 1\ea\right)\;.
 \la{Af}
\ee
These matrices  satisfy
$A_f^3=I$.

\item The jump matrix on the stem of the cherry attached to a vertex $v$ (which has valence  $2q+1$ on $\Sigma$) is chosen such that the total monodromy around $v$ is trivial  \eqref{nomono}.  

Let us assume  that all the edges are outgoing from $v$ using if necessary \eqref{edgereverse}. Then 
we deduce the following form of $M_v^0$ for each $v\in \mathbf V(\Sigma_0)$:
\be
M_v^0=\left(\prod_{f\prec e\perp v}^q  A_{f} S_e\right)^{-1}=\left(\ba{ccc} m_{v;1} & 0 & 0 \\ \star & {m_{v,2}}{m_{v;1}}^{-1} & 0\\
\star & \star & m_{v;2}^{-1} \ea\right)\;
\la{J03}
\ee
where 
\be
m_{v;1} =\left( \prod_{f\perp v} x_f \right)\left(\prod_{e\perp v} z_{e;1} z^2_{e;2}\right);  \qquad 
m_{v;2} =\left( \prod_{f\perp v} x^2_f \right)\left(\prod_{e\perp v}z^2_{e;1} z_{e;2}\right)\; ;  
 \ee
the 
change of orientation of some edge $e$ is equivalent to the interchange of $z_{e;1}$ and $z_{e;2}$.

\end{enumerate}

Since in the $SL(3)$ case there is only one face  variable $x_f = x_{f;111}$ for each face of $
\Sigma_0$  the  formula for $\omega_\M$ simplifies considerably since the  term $\omega_f$ vanishes. The expression (\ref{ototal}) takes the following form:
$$
{{2\Oh}}= \Omega(\Sigma_{_{FG}})=
\sum_{v\in V(T)} \sum_{e'\prec e \perp v} \sum_{j,k=1}^{2} \CM_{jk}  \d \zeta_{e;j} \wedge \d \zeta_{e';k}  
$$
$$
+3\sum_{v\in V(T)} \le( \sum_{e\prec f\perp v} 
 \d \xi_f \wedge (2 \d \zeta_{e;1} + \d\zeta_{e;2}) 
+\sum_{f\prec e\perp v}  
(2 \d \zeta_{e;1} + \d\zeta_{e;2})\ri) \wedge \d \xi_{f} 
$$
$$
 + 6 \sum_{f' \prec f\atop f, f'\perp v} \d \xi_f \wedge \d \xi_{f'}  +6\sum_{v\in V(T)} \sum_{j=1}^2  \d \rho_{v;j} \wedge  \d\mu_{v;j}
$$
where in the $SL(3)$ case we have  
$$
\CM= \le(\begin{array}{cc}
6 & 3\\
3 & 6 
\end{array}\ri).
$$
It is always understood that the edges are oriented away from $v$ and that $\zeta_{e;1} = \zeta_{-e;2}$ (i.e. under the change of orientation of an edge $e$ the variables 
$\zeta_{e;1}$ and $\zeta_{e;2}$ interchange).

{We notice that the $SL(3)$ Goldman bracket was studied before from various perspectives: in \cite{Kim} 
the Poisson tensor was expressed explicitly in the case of the three-punctured sphere in terms of trace coordinates.
In \cite{Lawton} the  symplectic form on a symplectic leaf was expressed in terms of the $SL(3)$ analogs of Fenchel-Nielsen coordinates.}


\end{document}